\documentclass[prl,
reprint,
 amsmath,amssymb,
 aps,
]{revtex4-2}

\usepackage{graphicx}
\usepackage{dcolumn}
\usepackage{bm}
\usepackage{physics}
\usepackage[colorlinks=true, citecolor=red, linkcolor=blue, urlcolor=black]{hyperref}
\usepackage{here}

\begin{document}


\title{Robust Spin-1 Haldane Topology under Itinerant Doping}

\author{Satoshi Nishimoto}
\email{s.nishimoto@ifw-dresden.de}
\affiliation{Department of Physics, Technical University Dresden, 01069 Dresden, Germany}
\affiliation{Institute for Theoretical Solid State Physics, IFW Dresden, 01069 Dresden, Germany}

\date{\today}

\begin{abstract}
	We study itinerant hole doping of the spin-$1$ Haldane
	symmetry-protected topological (SPT) phase in two $t$--$J$ chains
	coupled by ferromagnetic (FM) Hund exchange. Density-matrix
	renormalization group calculations show that string order,
	a finite spin gap, and the Haldane entanglement structure coexist
	over a broad doping range. Combined with entanglement-entropy
	scaling consistent with a single gapless charge mode, these
	results support a metallic doped Haldane SPT with a $C1S0$
	low-energy description. At low doping, short-range charge
	correlations reduce the density of spin-$1/2$ defect rungs,
	whereas at larger doping their positions become nearly
	uncorrelated while their spins are collectively incorporated
	into the correlated spin sector. The doped Haldane SPT terminates in
	a fully spin-polarized FM phase, which forms a reentrant pocket
	at high doping in the global phase diagram. Dilute-limit
	analyses trace the two FM boundaries to spin--charge
	factorization at weak Hund coupling and to rung-triplet binding
	with reduced pair mobility at strong Hund coupling. Our results
	show how mobile carriers reorganize the Haldane spin structure
	while preserving its characteristic signatures.
\end{abstract}

\maketitle


\textit{Introduction.}---
Itinerant doping poses a stringent test of the symmetry-protected
topological (SPT) Haldane phase of a spin-$1$ chain. The Haldane phase
is characterized by a finite spin gap, nonlocal string order,
fractionalized spin-$1/2$ edge degrees of freedom, and a characteristic
entanglement structure~\cite{Haldane1983a,Haldane1983b,AKLT1987,
	denNijs1989,Kennedy1992,Oshikawa1992,Pollmann2010,Pollmann2012}.
Several of these hallmarks have recently been observed directly in
ultracold Fermi--Hubbard ladders, where the Haldane phase remains
robust against substantial interaction-induced density
fluctuations~\cite{Sompet2022}. Static nonmagnetic dilution, by
contrast, interrupts the spin backbone and liberates effective
spin-$1/2$ moments~\cite{Hyman1997,Kenzelmann2003}. With itinerant
doping, however, the defects become mobile quantum degrees of freedom
whose spatial and spin correlations can reorganize.

Weak-coupling theory for Hund-coupled chains anticipated a conducting
spin-gapped descendant of the Haldane state~\cite{Fujimoto1995},
while related field-theoretical work established spin-gap formation
more generally in interacting double-chain systems~\cite{Nagaosa1996}.
Subsequent numerical studies found robust spin gaps and Haldane-related
correlations in doped spin-$1$ and ladder
systems~\cite{Ammon2000,Shirakawa2008,Laurell2024}. Topologically distinct Luttinger liquids and an anisotropy-driven transition between them were identified in a lightly doped ferromagnetic (FM)-rung ladder~\cite{Jiang2018}. What remains unclear is how broadly nonlocal string order, a bulk spin gap, and the characteristic Haldane entanglement structure coexist as doping increases. Equally important is how the spin sector accommodates a substantial density of mobile spin-1/2 defects, particularly when their short-range positional correlations become weak.

Here we address these questions in two $t$--$J$ chains coupled
by an FM Hund interaction on the rungs.
Using finite and infinite density-matrix renormalization group
(DMRG) calculations, we find that finite string order, a nonzero
bulk spin gap, and the characteristic Haldane entanglement
structure coexist over a broad doping range.
Entanglement-entropy scaling~\cite{Calabrese2004,Lauchli2008}
at representative points is consistent with a central charge $c\simeq1$.
These results point to a metallic doped Haldane SPT whose low-energy
theory contains one gapless charge mode and no gapless spin mode ($C1S0$),
placing it in the broader context of gapless SPT phases, in which protected
topological structure coexists with gapless bulk
modes~\cite{Keselman2015,Scaffidi2017,Keselman2018}.

The mechanism sustaining this state evolves from short-range charge
correlations that
reduce the density of spin-$1/2$ defect rungs at low doping to
collective incorporation of their spins into the correlated spin
sector at larger doping, where their positions become nearly
uncorrelated. At high doping, the global phase diagram also reveals a
reentrant FM pocket within the doped Haldane SPT, whose boundaries
originate from spin--charge factorization at weak Hund coupling and
the reduced mobility of rung-triplet pairs at strong Hund coupling.

\begin{figure}[bt]
	\centering
	\includegraphics[width=0.9\linewidth]{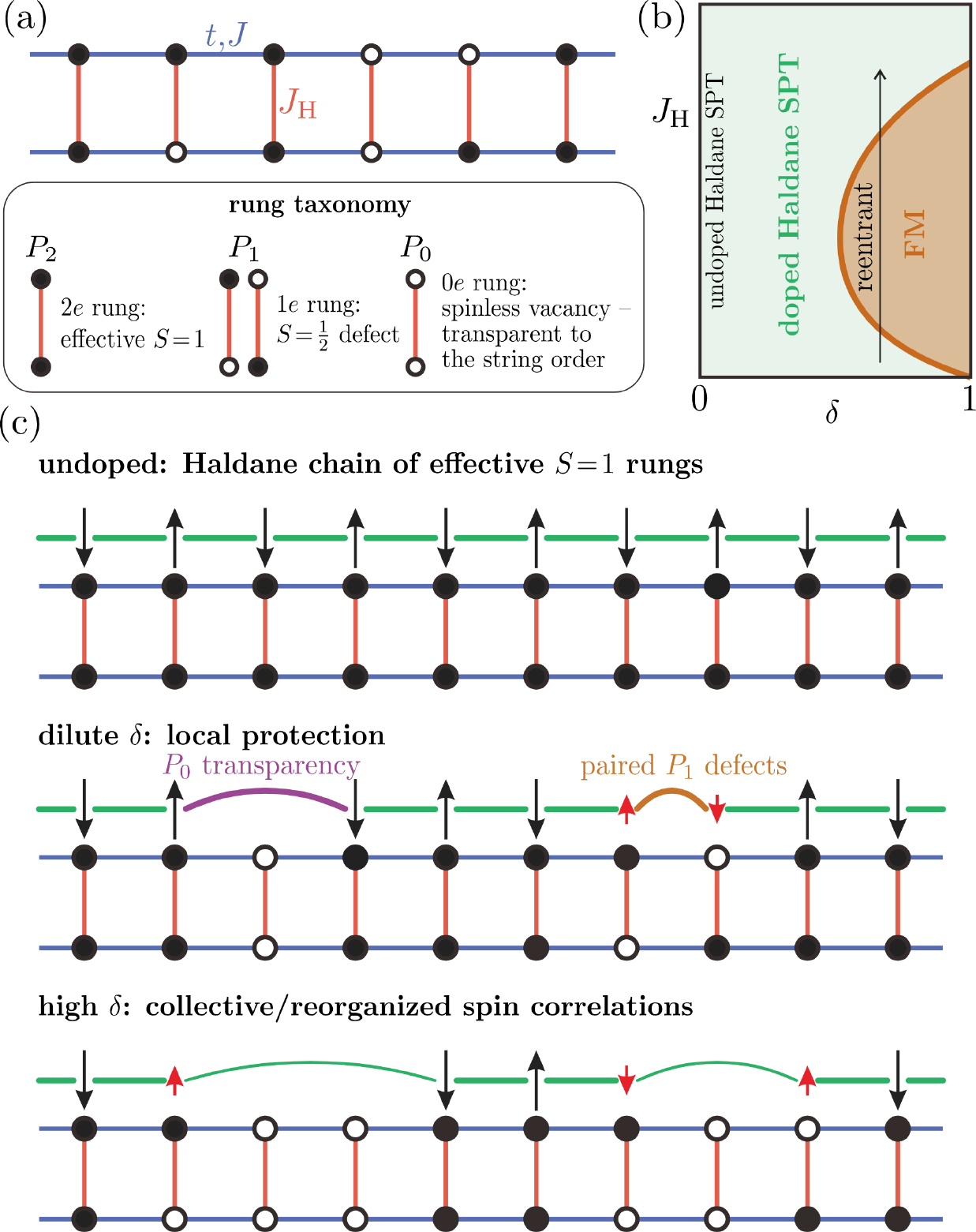}
	\caption{
		(a) Model and local rung taxonomy. Filled (open) circles denote
		electrons (holes).
		(b) Schematic evolution of the Haldane spin structure with doping.
		Black (red) arrows denote $S=1$ rung spins ($S=1/2$ defects), and
		green lines indicate the correlated spin backbone.
		(c) Schematic phase diagram in the $(J_{\rm H},\delta)$ plane.
	}
	\label{fig:lattice}
\end{figure}

\begin{flushleft}
	\begin{figure*}[t]
		\centering
		\includegraphics[width=0.85\textwidth]{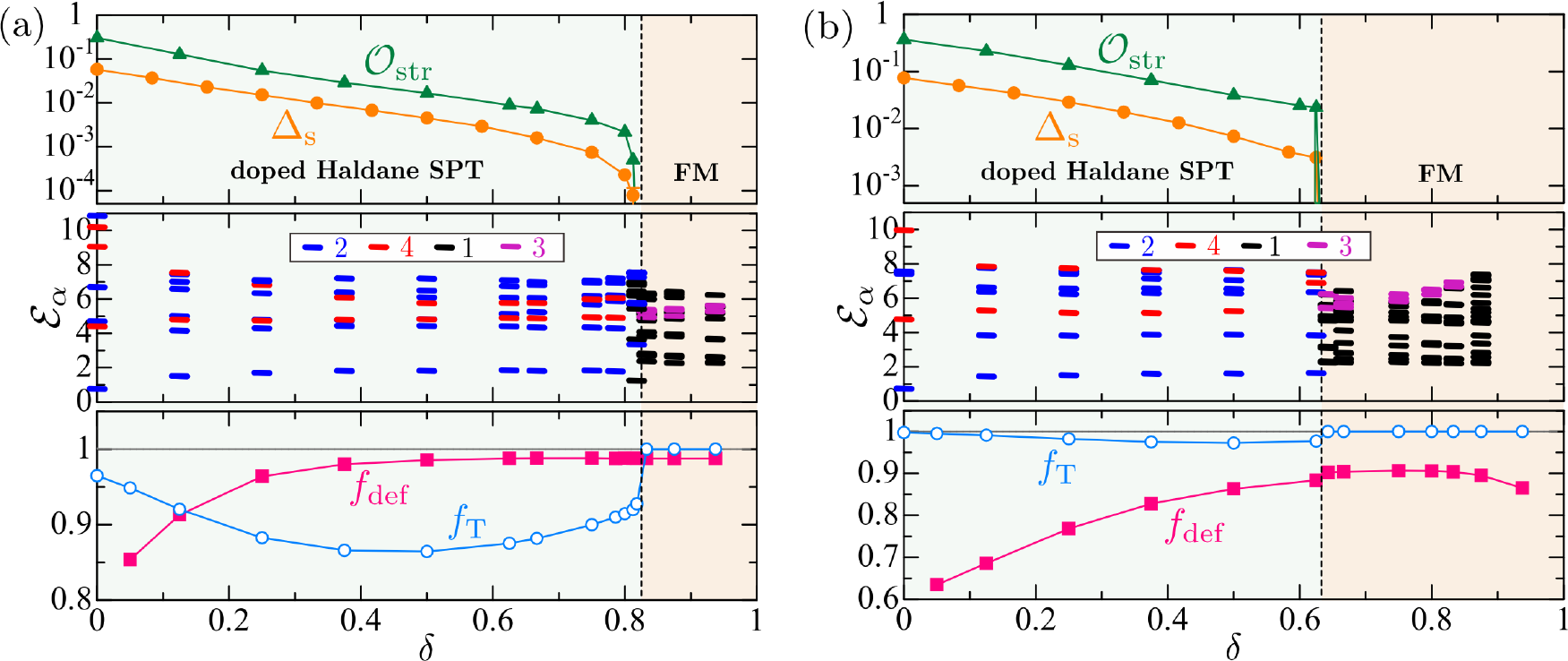}
		\caption{
			Doping evolution at $J=0.4$ for (a) $J_{\rm H}=0.5$ and
			(b) $J_{\rm H}=3.0$. Top: $\mathcal{O}_{\rm str}$ and
			$\Delta_s$. Middle: lowest entanglement levels
			$\mathcal{E}_\alpha$ at an inter-rung cut; colors denote
			multiplicity. Bottom: $f_{\rm def}$ and $f_{\rm T}$.
			Dashed lines mark $\delta_c$.
		}
		\label{fig:transition}
	\end{figure*}
\end{flushleft}

\textit{Model and diagnostics.}---
We consider two $t$--$J$ chains coupled by an FM Hund interaction on
the rungs [Fig.~\ref{fig:lattice}(a)],
\begin{equation}
	\begin{aligned}
		H={}&-t\sum_{i,a,\sigma}
		\left(
		\widetilde c_{i,a,\sigma}^{\dagger}
		\widetilde c_{i+1,a,\sigma}
		+\mathrm{H.c.}
		\right)
		\\
		&+J\sum_{i,a}
		\left(
		{\bf S}_{i,a}\cdot{\bf S}_{i+1,a}
		-\frac14 n_{i,a}n_{i+1,a}
		\right)
		\\
		&-J_{\rm H}\sum_i
		{\bf S}_{i,1}\cdot{\bf S}_{i,2}.
	\end{aligned}
	\label{eq:H}
\end{equation}
Here $i$ labels the rungs, $a=1,2$ the legs, and
$\widetilde c_{i,a,\sigma}
=c_{i,a,\sigma}(1-n_{i,a,\bar{\sigma}})$
excludes double occupancy. The two chains, with hopping $t$ and
antiferromagnetic (AFM) exchange $J$, are coupled only through the FM
Hund exchange $J_{\rm H}$; there is no interleg single-particle
hopping. We define the hole concentration as $\delta=1-n$, where
$n=\langle n_{i,a}\rangle$ is the electron density per site, and use
$t=1$ as the energy unit. At $\delta=0$, the FM Hund coupling aligns
the rung spins into effective $S=1$ moments, yielding the Haldane phase
for all $J>0$ and $J_{\rm H}>0$~\cite{paper1}.

The Hamiltonian preserves spin-rotation, time-reversal, and
bond-centered inversion symmetries, as well as leg-exchange symmetry; these include the symmetries protecting the Haldane phase~\cite{Pollmann2012}. At finite doping, however, the local Hilbert space contains both integer- and half-integer-spin states, so the persistence of the Haldane spin-sector structure is not guaranteed. We therefore assess it through the combined bulk, entanglement, and boundary diagnostics described below.

Upon doping, three local rung configurations become relevant
[Fig.~\ref{fig:lattice}(a)]: a two-electron rung, whose spin state is
favored to be triplet by $J_{\rm H}$, a singly occupied rung carrying
spin $1/2$, and an empty spinless rung, with probabilities
$P_2=\langle n_{i,1}n_{i,2}\rangle$, $P_1=2(n-P_2)$, and
$P_0=1-2n+P_2$, respectively, by leg symmetry. We quantify the
suppression of spin-$1/2$ defect rungs by
$f_{\rm def}=P_1/[2n(1-n)]$, for which $f_{\rm def}=1$ corresponds to
statistically independent occupations of the two legs at fixed
density $n$. The conditional triplet fraction of the two-electron
rungs is $f_{\rm T}=T/P_2$, with
$T=\langle{\bf S}_{i,1}\cdot{\bf S}_{i,2}
+\tfrac34n_{i,1}n_{i,2}\rangle$, where the operator inside the
expectation value projects onto the rung-triplet sector.
These local quantities characterize the rung composition but are not,
by themselves, diagnostics of topological order.

We characterize the doped Haldane SPT using three complementary
spin-sector diagnostics. Defining the total rung spin
$\widetilde{\bf S}_i={\bf S}_{i,1}+{\bf S}_{i,2}$, the string-order
parameter (SOP) is
$\mathcal{O}_{\rm str}
=-\lim_{r\rightarrow\infty}
\langle\widetilde S_i^z
\exp(i\pi\sum_{k=i+1}^{i+r-1}\widetilde S_k^z)
\widetilde S_{i+r}^z\rangle$
~\cite{denNijs1989,Kennedy1992}. The spin gap at fixed electron number
is
$\Delta_s=E_0(S_{\mathrm{tot}}^z=1)
-E_0(S_{\mathrm{tot}}^z=0)$.
Because fractionalized Haldane edge spins can produce a near-zero
excitation on an open ladder, we examine the spatial profile of the
excess magnetization to distinguish boundary-localized and bulk
excitations. Boundary-localized spin excitations persist at
representative finite dopings within the doped Haldane SPT, providing
a complementary boundary manifestation of the surviving Haldane spin
sector. To extract the thermodynamic bulk spin gap, we modify the
boundary environment to suppress these contributions and perform a
finite-size extrapolation, as detailed in the Supplemental Material~\cite{SM}.

Finally, denoting the Schmidt coefficients across an inter-rung cut
by $\lambda_\alpha$, we define the entanglement spectrum (ES) as
$\mathcal{E}_\alpha=-\ln\lambda_\alpha^2$~\cite{Li2008}. 
At each inter-rung cut, we analyze the spin-resolved low-lying
ES within the dominant particle-number sector, defined as the
sector containing the largest Schmidt coefficient.
The systematic organization of these levels into
nearly degenerate even multiplets provides an independent probe of the
Haldane entanglement structure~\cite{Pollmann2010,Pollmann2012}. 
Entanglement-entropy scaling~\cite{Calabrese2004,Lauchli2008} gives
estimates consistent with $c\simeq1$ at representative points and, 
given the finite spin gap, favors a $C1S0$ description.
Details of the ES and central-charge analyses are given
in the Supplemental Material~\cite{SM}.

Ground states are obtained using finite DMRG~\cite{White1992} and infinite DMRG
(iDMRG)~\cite{McCulloch2008}, conserving the total particle number and
$S_{\mathrm{tot}}^z$, with bond dimensions up to $\chi=10000$. Phase
boundaries toward the fully polarized state are located from the onset
of degeneracy between the lowest $S_{\mathrm{tot}}^z=0$ and fully polarized
energies at fixed filling; finite DMRG and iDMRG agree within the resolution of
the parameter grid. Further numerical details are given in the
Supplemental Material~\cite{SM}.

\textit{Doping robustness and transition.}---
Figure~\ref{fig:transition} shows the doping evolution for two
representative Hund couplings, $J_{\rm H}=0.5$ and $3.0$, at fixed
$J=0.4$. For $J_{\rm H}=0.5$, the SOP
$\mathcal{O}_{\rm str}$ remains finite over the entire
nonferromagnetic range up to $\delta_c\simeq0.82$, while the spin gap
$\Delta_s$ is finite over a broad doping range but becomes very small
on approaching $\delta_c$; correspondingly, the ideal even-fold ES
pattern is partially lifted near $\delta_c$, although pronounced
near-degenerate doublets remain on the Haldane side. For $J_{\rm H}=3.0$, both $\mathcal{O}_{\rm str}$ and $\Delta_s$
remain clearly finite up to $\delta=0.625$, immediately below
$\delta_c\simeq0.63$, and the low-lying ES levels are organized into
well-developed even-fold near-degenerate multiplets throughout.
In both cases, the doped Haldane SPT terminates at the transition to
the fully spin-polarized FM phase; its Haldane signatures remain well
developed immediately below the boundary for $J_{\rm H}=3.0$ but are
already strongly suppressed for $J_{\rm H}=0.5$.

The lower panels of Fig.~\ref{fig:transition} show relatively modest
changes in $f_{\rm def}$ and $f_{\rm T}$ across the transition.
Within the doped Haldane SPT, $f_{\rm def}$ increases toward unity
with doping, while $f_{\rm T}$ remains sizable.
The loss of the Haldane signatures despite these modest local
changes highlights the role of collective spin correlations.

In a weak-coupling description, this robustness can be understood
in terms of the total and relative charge modes and two spin
modes~\cite{Fujimoto1995}. Doping couples directly
to the total-charge field and renders the half-filled Umklapp
interaction incommensurate, leaving the total-charge mode gapless,
whereas Hund-generated interactions can continue to pin the
relative-charge and spin fields. Schematically,
\begin{equation}
	H_\mu
	\propto
	\mu\int dx\,
	\partial_x\phi_{\rho+},
	\qquad
	\mathcal{O}_{\rm str}
	\sim
	\left|
	\left\langle
	\cos(\beta_s\phi_{\sigma+})
	\right\rangle
	\right|^2.
	\label{eq:bosonization}
\end{equation}
Here $\beta_s$ is a convention-dependent coefficient. The
bosonization analysis is detailed in the Supplemental Material~\cite{SM}.

\begin{figure}[tb]
	\centering
	\includegraphics[width=0.9\linewidth]{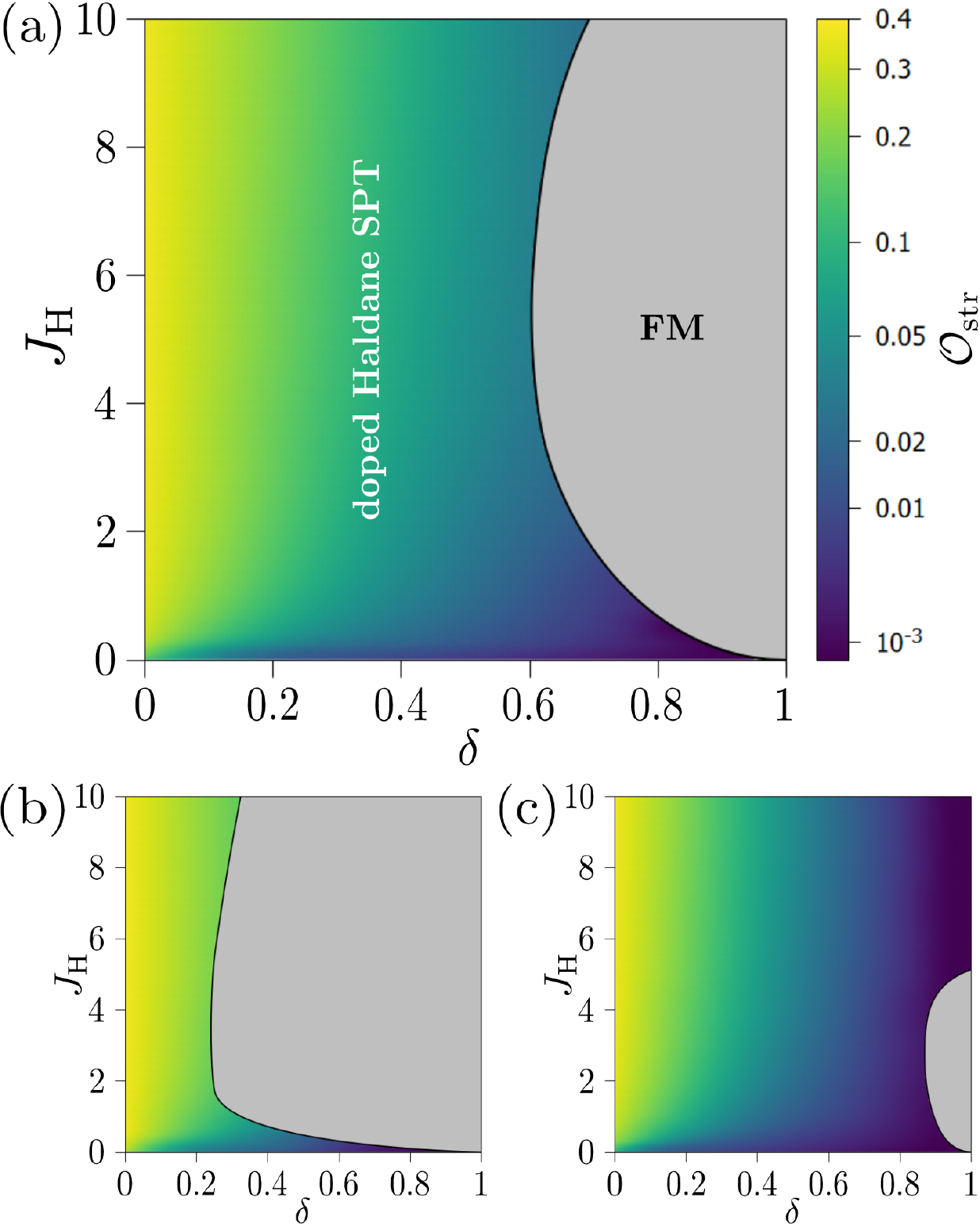}
	\caption{
		Phase diagram in the $(J_{\rm H},\delta)$ plane for
		(a) $J=0.4$, (b) $J=0.1$, and (c) $J=1.0$. The color scale
		shows the SOP $\mathcal{O}_{\rm str}$ in the doped Haldane
		SPT, while the gray regions denote the FM phase.
	}
	\label{fig:pd}
\end{figure}

\textit{Global phase diagram.}---
Figure~\ref{fig:lattice}(b) summarizes schematically the global phase
structure, and Fig.~\ref{fig:pd} shows the corresponding numerical
phase diagrams in the $(J_{\rm H},\delta)$ plane. For $J=0.4$
[Fig.~\ref{fig:pd}(a)], the doped Haldane SPT occupies a broad region,
with finite string order surviving to large hole concentrations over
a wide range of Hund couplings. Ferromagnetism appears only at
sufficiently high doping, where it cuts into the doped Haldane SPT,
consistent with the termination points in Fig.~\ref{fig:transition}.

The phase-diagram topology depends strongly on the AFM leg exchange
$J$ [Figs.~\ref{fig:pd}(b) and \ref{fig:pd}(c)]. Reducing $J$
enlarges the FM region, whereas increasing $J$ suppresses it. For
$J=1.0$ [Fig.~\ref{fig:pd}(c)], ferromagnetism is confined to a
high-doping pocket at intermediate Hund coupling. The system
therefore enters and leaves the FM phase upon increasing $J_{\rm H}$
at fixed high doping, producing a reentrant FM pocket within the
doped Haldane SPT.

In the dilute-electron limit, $n=1-\delta\rightarrow0$, the two FM
boundaries have complementary origins. On the weak-Hund side, charge
motion approaches the factorized limit~\cite{Ogata1990,Kruis2004}, where the
leading kinetic energy becomes insensitive to the spin configuration.
The squeezed-chain exchange is $J_b=JC_{nn}(n)/n$, with
$C_{nn}(n)=n^2-[\sin(\pi n)/\pi]^2$. Balancing the resulting AFM
energy against the leading Hund-energy gain of the FM state gives
the leading-order estimate
$J_{\rm H}^{c,-}(n)
=8\ln2\,J[1-(\sin(\pi n)/\pi n)^2]
\simeq18.24\,Jn^2$.
Thus, the lower FM boundary approaches $J_{\rm H}=0$ as
$\delta\rightarrow1$ and shifts toward larger $J_{\rm H}$ with
increasing $J$, consistent with Fig.~\ref{fig:pd}.

The strong-Hund boundary has a distinct dilute-limit origin. Strong
$J_{\rm H}$ binds electrons on the two legs into rung-triplet pairs,
and the minimal four-electron problem probes the effective interaction
between two such pairs. For $N=4$ electrons on a ladder of length $L$,
we define
$\Delta E_4=
E_0(L,N=4,S_{\mathrm{tot}}^z=2)
-E_0(L,N=4,S_{\mathrm{tot}}^z=0)$.
This quantity vanishes in the fully polarized region because the two
states belong to the same $S_{\mathrm{tot}}=2$ multiplet and becomes
positive on the non-fully-polarized side. Extrapolating the resulting
finite-size thresholds to $L\rightarrow\infty$ gives
$J_{\rm H}^{c,+}(0)/t\simeq27.45$, $12.80$, and $5.14$ for
$J/t=0.1$, $0.4$, and $1.0$, respectively. These values explain the
evolution in Fig.~\ref{fig:pd}: the upper boundary lies beyond the
displayed $J_{\rm H}$ range for $J=0.1$ and $0.4$, whereas for
$J=1.0$ it occurs within the displayed range and closes the FM
pocket. Details of the dilute-limit analysis are
given in the Supplemental Material~\cite{SM}.

At strong $J_{\rm H}$, the dominant rung-triplet binding is shared
by the competing spin sectors, while the pair
mobility decreases as $t_T\simeq8t^2/J_{\rm H}$ and the projected AFM
scale remains $J_{\rm eff}=J/2$. The two limits thus explain the
intermediate Hund-coupling window of ferromagnetism.

\begin{figure}[tb]
	\centering
	\includegraphics[width=0.9\linewidth]{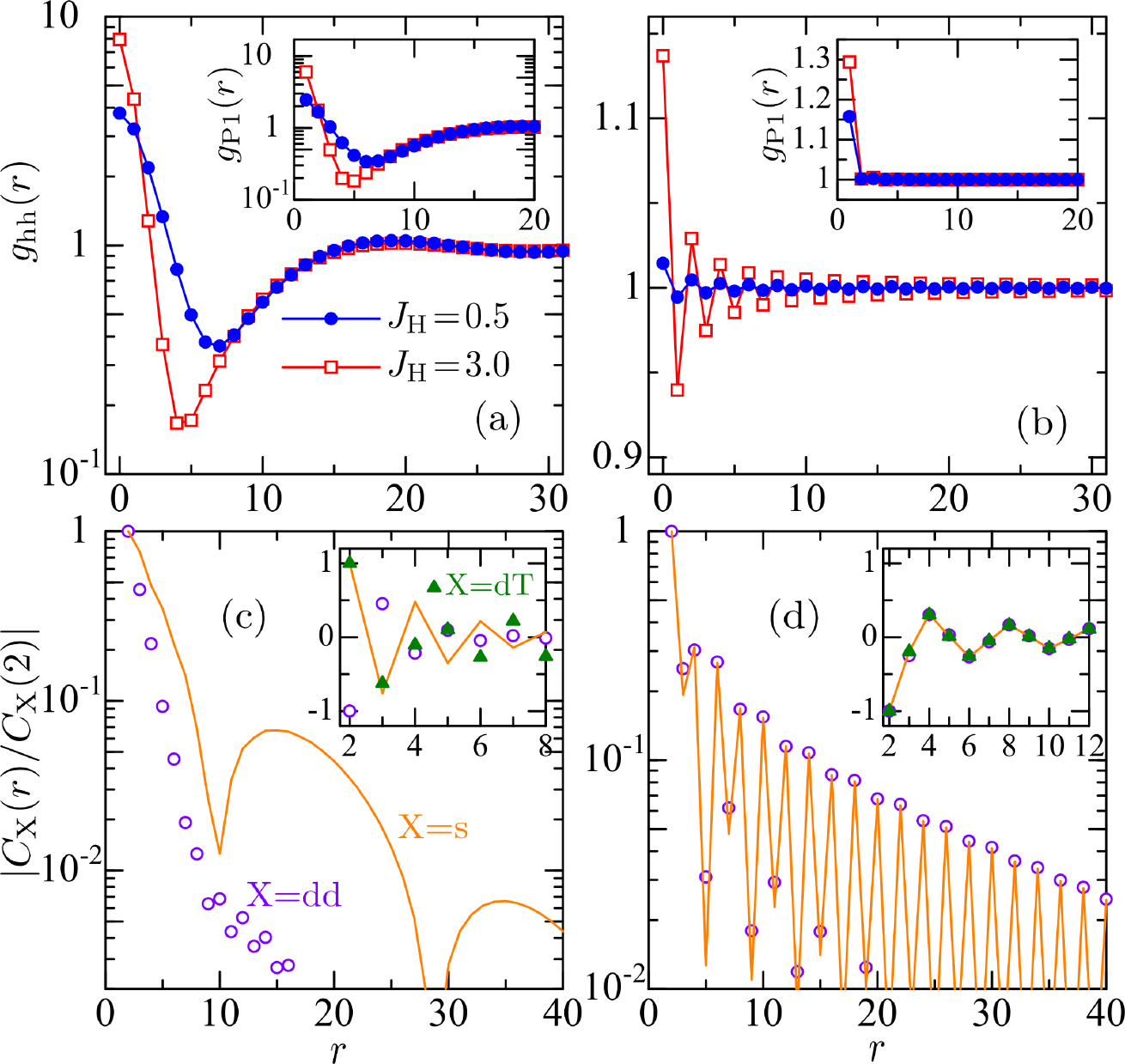}
	\caption{
		Hole and defect-resolved correlations at $J=0.4$.
		(a,b) $g_{hh}(r)$ at $\delta=0.05$ and $0.5$, respectively,
		for $J_{\rm H}=0.5$ and $3.0$; the insets show $g_{P_1}(r)$.
		(c,d) $|C_X(r)|/|C_X(2)|$ for $X\in\{s,dd\}$ at
		$J_{\rm H}=0.5$ for the same respective dopings; the insets
		show $C_X(r)/|C_X(2)|$ for $X\in\{s,dd,dT\}$.
	}
	\label{fig:mechanism}
\end{figure}

\textit{Microscopic mechanism.}---
Figure~\ref{fig:lattice}(c) schematically illustrates two ways in
which the Haldane spin sector can accommodate mobile defects. At
dilute doping, short-range charge correlations reduce the density of
spin-$1/2$ defect rungs and favor locally less disruptive
arrangements, consistent with schematic motifs such as
$P_2$--$P_0$--$P_2$ and $P_2$--$P_1$--$P_1$--$P_2$.
An empty rung contributes unity to the string phase factor,
whereas a singly occupied spin-$1/2$ rung contributes $\pm i$.
Although no direct exchange bond spans an empty rung, hopping
changes its occupancy, so it does not constitute a permanent cut
as a static vacancy would.
At larger doping, these positional correlations weaken, raising
the question of whether a finite density of spin-$1/2$ defects
can instead be accommodated collectively by the spin sector.
Figure~\ref{fig:mechanism} tests this question directly.

We first calculate the normalized interleg hole correlation
\begin{equation}
	g_{hh}(r)
	=
	\frac{
		\overline{\left\langle
			n^{h}_{i,1}n^{h}_{i+r,2}
			\right\rangle}
	}{
		\overline{
			\left\langle n^{h}_{i,1}\right\rangle
			\left\langle n^{h}_{i+r,2}\right\rangle}
	},
	\qquad
	n^h_{i,a}=1-n_{i,a},
	\label{eq:ghh}
\end{equation}
where the bar denotes an average over rungs. We also calculate
the analogous normalized correlation $g_{P_1}(r)$ for $r\geq1$,
using the single-occupancy projector
$\hat p_{1,i}=n_{i,1}+n_{i,2}-2n_{i,1}n_{i,2}$.

At dilute doping [Fig.~\ref{fig:mechanism}(a)], the holes show strong
short-range correlations, with $g_{hh}(0)=3.7$ and $7.9$ for
$J_{\rm H}=0.5$ and $3.0$, respectively, followed by a depletion
region and recovery toward the uncorrelated value $g_{hh}=1$. Since
$g_{hh}(0)=P_0/\delta^2$ exactly, the rung relations above give
$f_{\rm def}=1-[g_{hh}(0)-1]\,\delta/n$, so that the enhanced
probability of empty rungs directly reduces the density of
spin-$1/2$ defect rungs. At $\delta=0.05$, this yields
$f_{\rm def}\simeq0.85$ and $0.63$ for $J_{\rm H}=0.5$ and $3.0$,
respectively. This redistribution therefore provides a natural local mechanism for suppressing potentially disruptive $P_1$ defects.
The nearest-neighbor enhancement of $g_{P_1}(r)$
[inset of Fig.~\ref{fig:mechanism}(a)] supports an increased
probability of adjacent singly occupied rungs, consistent with
the schematic motifs in Fig.~\ref{fig:lattice}(c).

At larger doping [Fig.~\ref{fig:mechanism}(b), $\delta=0.5$], these
short-range positional correlations are substantially weakened. The
on-rung enhancement is reduced to $g_{hh}(0)\simeq1.01$ and $1.14$
for $J_{\rm H}=0.5$ and $3.0$, respectively, corresponding to
$f_{\rm def}\simeq0.99$ and $0.86$. For $J_{\rm H}=0.5$, this gives $P_1\simeq0.495$, so singly occupied rungs constitute nearly half
of all rungs. Away from $r=0$, both
$g_{hh}(r)$ and $g_{P_1}(r)$ rapidly approach their uncorrelated
values. Thus, despite a residual on-rung correlation at stronger Hund
coupling, the defect positions exhibit little correlation at finite
separation. Nevertheless, the doped Haldane SPT survives at the same doping
[Fig.~\ref{fig:transition}], showing that local defect avoidance alone
can no longer account for its stability.

For the spin-correlation analysis, we focus on $J_{\rm H}=0.5$,
where $f_{\rm def}\simeq1$ at $\delta=0.5$.
We decompose the total rung spin into contributions from singly
occupied and two-electron rungs,
$\widetilde S_i^z=d_i^z+T_i^z$, with
$d_i^z=S^z_{i,1}(1-n_{i,2})+S^z_{i,2}(1-n_{i,1})$ and
$T_i^z=S^z_{i,1}n_{i,2}+S^z_{i,2}n_{i,1}$, and define the connected
correlations
\begin{equation}
	C_{XY}(r)
	=
	\overline{
		\left\langle X_i^zY_{i+r}^z\right\rangle
		-
		\left\langle X_i^z\right\rangle
		\left\langle Y_{i+r}^z\right\rangle
	},
	\qquad
	X,Y\in\{d,T\}.
	\label{eq:CXY}
\end{equation}
The total rung-spin correlation satisfies
$C_s(r)=C_{dd}(r)+C_{dT}(r)+C_{Td}(r)+C_{TT}(r)$. 
The plotted channels are independently normalized by
$|C_X(2)|$ to compare their spatial profiles.

At dilute doping [Fig.~\ref{fig:mechanism}(c)], the normalized
defect--defect correlation $C_{dd}(r)$ decays more rapidly than
the total correlation $C_s(r)$ and does not reproduce its
pronounced long-distance modulation. Because $C_{dd}(r)$ is
weighted by the joint probability of finding singly occupied
rungs at both endpoints, its distance dependence contains both
positional and spin correlations, even after normalization
at $r=2$. The difference in spatial structure persists after conditioning
on single occupancy at both endpoints; see the Supplemental
Material~\cite{SM}.

A qualitatively different behavior appears at $\delta=0.5$
[Fig.~\ref{fig:mechanism}(d)]. Although the defect positions exhibit
little correlation at finite separation, the normalized $C_{dd}(r)$
closely follows the normalized $C_s(r)$ over the displayed range,
sharing its oscillatory structure and decay envelope. 
The normalized mixed defect--background channel $C_{dT}(r)$
also shares this spatial structure. Together with the finite bulk
spin gap, this shared structure indicates that the defect spins
participate in the same correlated spin sector rather than behaving
as nearly free moments. The contrast with dilute doping supports
a crossover from local defect avoidance to collective spin incorporation.

\textit{Conclusion.}---
We have shown that finite string order, a finite bulk spin gap,
and the characteristic Haldane entanglement structure coexist
over a broad range of hole doping.
Together with entanglement-entropy scaling consistent with
$c\simeq1$ at representative points, these results support
a metallic doped Haldane SPT with a $C1S0$ low-energy description.
The way mobile defects are accommodated evolves from local
defect avoidance at low doping to collective incorporation
of their spins into the correlated spin sector at larger doping.
The global phase diagram further exhibits a reentrant FM pocket
within the doped Haldane SPT at high doping.
The pairing properties of this regime~\cite{Feiguin2008,Laurell2024}
lie outside the present scope and will be addressed separately.
Fermi--Hubbard ladders with site-resolved detection have already
realized the Haldane phase~\cite{Sompet2022}.
Since the rung probabilities $P_m$, $f_{\rm def}$, and $g_{hh}(r)$
can be obtained from charge snapshots, while the string and
defect-resolved spin correlations can be constructed from
spin-resolved snapshots, controlled doping of such platforms
would allow experimental tests of the predicted Haldane
signatures and the crossover identified here.

{\it Acknowledgments.}---
We thank Ulrike Nitzsche for technical assistance. This project is funded by the German Research Foundation (DFG) via the projects A05 of the Collaborative Research Center SFB 1143 (project-id 247310070). The iDMRG calculations in this work were performed using the TeNPy tensor-network library \cite{Hauschild2018}.

\bibliography{dopedHaldane}

\end{document}



\begin{center}
	\large
	{\bf Supplemental Material for\\ ``Robust Spin-1 Haldane Topology under Itinerant Doping''}
	
	\normalsize
	\vspace{3.0mm}
	Satoshi Nishimoto
	
	\small
	
	\vspace{1.5mm}
	{\it
		Department of Physics, Technical University Dresden, 01069 Dresden, Germany\\
		Institute for Theoretical Solid State Physics, IFW Dresden, 01069 Dresden, Germany
	}
\end{center}

\begin{abstract}
\end{abstract}
\maketitle

\vspace{-1em}
\tableofcontents
\vspace{1em}
\hrule
\vspace{1.5em}

\newpage

\section{Numerical methods and convergence}
\label{sec:numerics}

We investigate the ground-state properties of the doped two-leg $t$--$J$ ladder using both finite-system density-matrix renormalization group (DMRG) and infinite DMRG (iDMRG) calculations. The leg hopping is taken as the unit of energy, $t=1$. In all calculations, we conserve the total particle number and the total spin projection $S_{\mathrm{tot}}^{z}$. The MPS bond dimension is denoted by $\chi$.

Finite-system DMRG calculations with open boundary conditions are used to evaluate ground-state energies, spin gaps, von Neumann entanglement entropies, selected local observables, and the phase boundary toward the fully polarized state. We retain up to $\chi=10000$ states. For the most demanding finite-system calculations, the maximum discarded weight in the final sweeps at the largest bond dimension is of order $10^{-8}$. The stability of the relevant observables against increasing $\chi$ was checked explicitly, as documented in the respective sections below.

For the finite-size scaling of the spin gap, we typically use seven system sizes spanning $L=24$ to $L=168$. For some fillings, the precise values of $L$ are adjusted slightly so that the target density can be represented exactly by an integer particle number. Here and below, $L$ denotes the number of rungs. The definition of the finite-size spin gap, the boundary geometry, and the extrapolation to the thermodynamic limit are described in Sec.~\ref{sec:spin_gap_boundary}.

Bulk quantities, including the string-order parameter, the entanglement spectrum, correlation functions, and local rung-state probabilities, are evaluated primarily using iDMRG. The iDMRG unit cell is chosen to represent the target density exactly. We use unit cells containing up to 22 rungs, corresponding to 44 MPS sites. The largest unit cell is required, in particular, to access the hole concentration $\delta=36/44$ close to a phase boundary for $J=0.4$ and $J_{\rm H}=0.5$. Most iDMRG results are obtained with bond dimensions up to $\chi=6000$. For the
22-rung unit cell, the calculations were extended to $\chi=10000$; the relevant observables were already nearly converged at $\chi=6000$.

Because the doped metallic state contains a gapless charge mode, the total MPS correlation length and the entanglement entropy need not saturate with increasing $\chi$. We therefore assess convergence directly for each observable relevant to the doped Haldane regime. For the entanglement-spectrum analysis, calculations are performed over a sequence of increasing bond dimensions, using the converged state at the preceding bond dimension as the initial state. The quantum-number-sector selection, averaging over inequivalent inter-rung cuts, and convergence of the lowest entanglement-level splitting are discussed in Sec.~\ref{sec:entanglement_spectrum}. The discarded-weight extrapolation of the long-distance string-order parameter is presented in Sec.~\ref{sec:string_order}.

The local rung-state probabilities were calculated independently using finite-system DMRG and iDMRG and were found to be mutually consistent. The phase boundaries toward the fully polarized state were also determined independently using the two approaches. In finite-system DMRG, we locate the onset of degeneracy between the lowest energies in the $S_{\mathrm{tot}}^{z}=0$ and $S_{\mathrm{tot}}^{z}=S_{\max}$ sectors at fixed particle number, where $S_{\max}=N/2$. In iDMRG, we separately optimize the $S_{\mathrm{tot}}^{z}=0$ state and the fully polarized state at the same particle density. The latter is obtained by taking all electrons to have the same spin orientation, so that the spin projection per unit cell is maximal. The corresponding boundary is located from the degeneracy of the optimized energy densities of the two sectors. The ferromagnetic phase boundaries obtained from finite-system DMRG and iDMRG agree within the resolution of the parameter grid used in the phase diagrams.

\section{Spin-gap calculation and boundary treatment}
\label{sec:spin_gap_boundary}

For a finite ladder with open boundary conditions, we define the spin gap at fixed electron number $N$ as
\begin{equation}
	\Delta_s(L,N)
	=
	E_0(L,N,S_{\mathrm{tot}}^z=1)
	-
	E_0(L,N,S_{\mathrm{tot}}^z=0),
	\label{eq:spin_gap}
\end{equation}
where $E_0(L,N,S_{\mathrm{tot}}^z)$ denotes the lowest energy in the sector with total spin projection $S_{\mathrm{tot}}^z$. A direct application of Eq.~(\ref{eq:spin_gap}) to an ordinary open ladder, however, does not generally yield the bulk spin gap. In the Haldane regime, the two open boundaries support fractionalized spin-$1/2$ degrees of freedom, whose finite-size splitting becomes exponentially small with increasing system size. Consequently, the lowest state in the $S_{\mathrm{tot}}^z=1$ sector can represent an edge excitation rather than a bulk spin excitation.

To distinguish these possibilities, we examine the rung-resolved excess magnetization
\begin{equation}
	\delta m_i
	=
	\left\langle S_i^z\right\rangle_{S_{\mathrm{tot}}^z=1}
	-
	\left\langle S_i^z\right\rangle_{S_{\mathrm{tot}}^z=0},
	\qquad
	S_i^z=S_{i,1}^z+S_{i,2}^z,
	\label{eq:delta_m}
\end{equation}
where the two expectation values are calculated at the same electron number. This quantity satisfies
\begin{equation}
	\sum_i\delta m_i=1.
	\label{eq:delta_m_sum}
\end{equation}
An edge excitation produces a profile localized near one of the two boundaries, whereas a bulk spin excitation has a spatially extended profile. Because the two boundary modes are asymptotically degenerate, the numerically obtained excitation may localize at either end of the ladder; the choice of end has no physical significance.

\begin{figure}[t]
	\centering
	\includegraphics[width=0.7\linewidth]{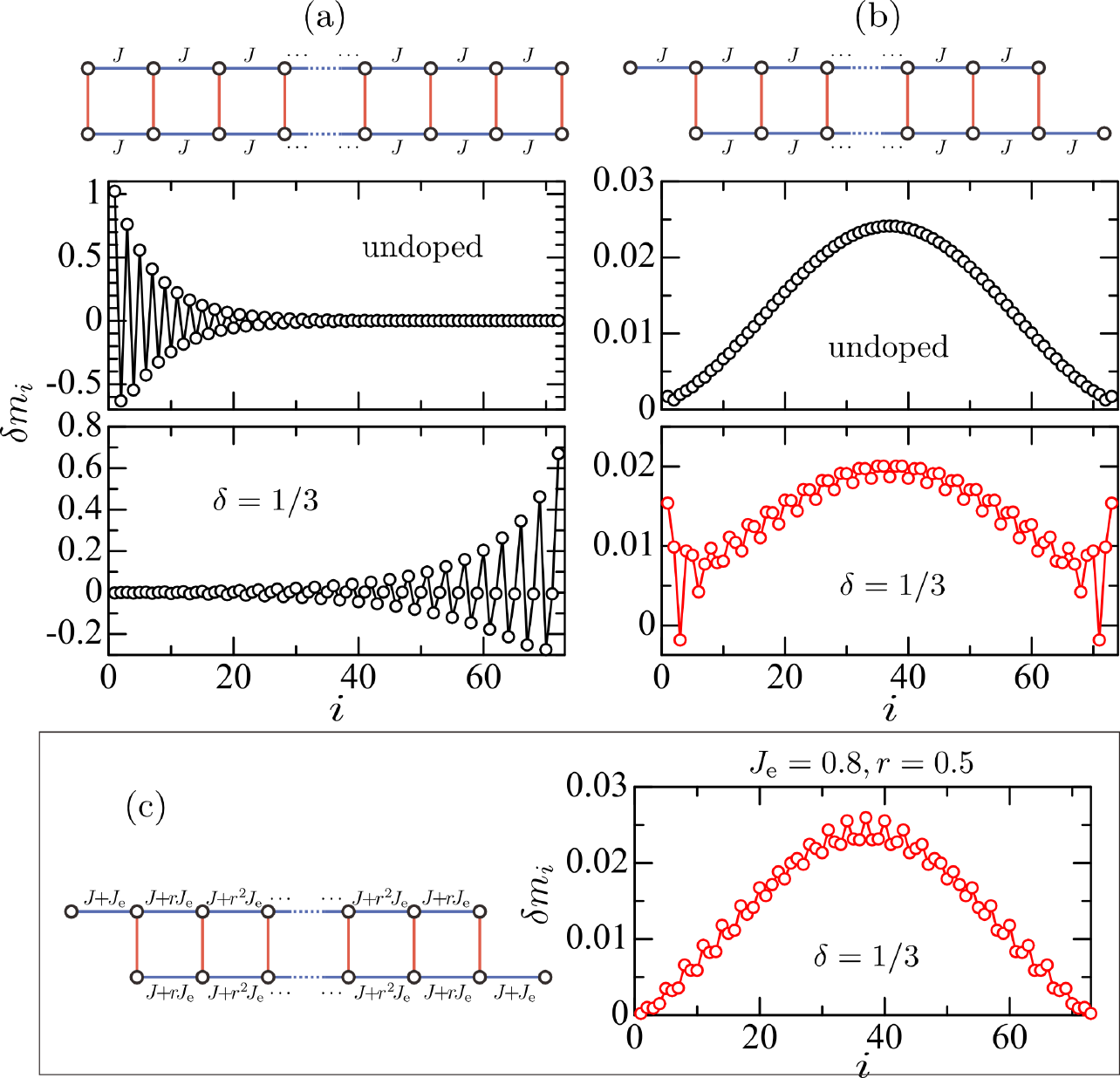}
	\caption{
		Rung-resolved excess magnetization $\delta m_i$ defined in
		Eq.~(\ref{eq:delta_m}) at $J=0.4$ and $J_{\mathrm H}=3.0$.
		(a) Ordinary open ladder with $L=72$ rungs (144 sites)
		at $\delta=0$ and $1/3$.
		(b) Modified open ladder with $L=73$ rung positions including 144 
		sites, obtained by removing one site from each terminal rung on
		opposite legs.
		(c) Results at $\delta=1/3$ after introducing the smoothly tapered
		edge exchange with $J_e=0.8$ and $r=0.5$.
		The upper diagrams illustrate the corresponding boundary geometries
		and exchange profiles.
	}
	\label{fig:spin_boundary}
\end{figure}

Figure~\ref{fig:spin_boundary}(a) shows $\delta m_i$ for an ordinary open ladder at $J=0.4$ and $J_{\mathrm H}=3.0$. At $\delta=0$, the strongly edge-localized and staggered profile is the familiar manifestation of the fractionalized Haldane edge spin. Importantly, an edge-localized excitation is also found at $\delta=1/3$. Its persistence upon doping provides a complementary boundary manifestation of the surviving Haldane-type spin structure in the doped Haldane SPT regime. This observation is consistent with the simultaneous survival of the string order, spin gap, and characteristic entanglement-spectrum structure discussed in the main text. We do not use this boundary signature by itself to identify the doped Haldane SPT.

To suppress the edge contribution when calculating the bulk spin gap, we use the modified open geometry illustrated in Fig.~\ref{fig:spin_boundary}(b). The ordinary ladder in Fig.~\ref{fig:spin_boundary}(a) has $L=72$ rungs and hence 144 sites. For the modified geometry, we take $L=73$ rung positions and remove one site from each terminal rung, choosing the removed sites on opposite legs. The resulting ladder also contains 144 sites. The two geometries can therefore be compared at the same electron number and hole concentration without changing the total number of physical sites.

More generally, for a modified ladder containing $L$ rung positions, the number of physical sites and the hole concentration are
\begin{equation}
	N_{\mathrm s}=2L-2,
	\qquad
	\delta=1-\frac{N}{N_{\mathrm s}}.
	\label{eq:modified_geometry_size}
\end{equation}
At $\delta=0$, the physical spin-$1/2$ remaining on each single-site terminal rung couples to and screens the corresponding fractionalized Haldane edge spin. This modification eliminates the near-zero edge excitation and produces the smooth, spatially extended profile shown in the upper panel of Fig.~\ref{fig:spin_boundary}(b).

In the doped system, however, removing the two terminal sites is not always sufficient. Mobile holes rearrange the local charge and spin environment near an open boundary and can weaken the screening provided by the remaining terminal site or shift the effective spin boundary toward the interior. The lowest excitation may then retain a residual boundary contribution, as illustrated by the oscillatory end structure superposed on the extended profile at $\delta=1/3$ in the lower panel of Fig.~\ref{fig:spin_boundary}(b). Such a contribution depends on the boundary environment and must be suppressed before Eq.~(\ref{eq:spin_gap}) can be interpreted as a bulk spin gap.

Practically, for parameter regions in which this residual boundary contribution occurs, we gradually enhance the antiferromagnetic leg exchange near both ends. Denoting the additional edge exchange by $J_e$, the modified leg couplings are chosen as
\begin{equation}
	J_{\mathrm{leg}}^{(k)}
	=
	J+r^kJ_e,
	\qquad
	k=0,1,2,\ldots,
	\qquad
	0<r<1,
	\label{eq:tapered_edge}
\end{equation}
where $k$ counts successive bond shells inward from each boundary. The two leg bonds directly connected to the single-site terminal rungs have coupling $J+J_e$. The next four leg bonds, consisting of two bonds near each end, have coupling $J+rJ_e$, the following four have coupling $J+r^2J_e$, and so forth. The modification is terminated once $r^kJ_e<0.1t$. This smooth tapering avoids introducing a sharp interface between the boundary-modified and unmodified regions.

The boundary profile and the resulting $\delta m_i$ for $J_e=0.8$ and $r=0.5$ are shown in Fig.~\ref{fig:spin_boundary}(c). The residual end structure visible in Fig.~\ref{fig:spin_boundary}(b) is suppressed, and the excess magnetization forms a smooth profile extending over the bulk. We therefore use the corresponding excitation energy in the finite-size extrapolation of the bulk spin gap. The boundary modification serves only to remove localized boundary excitations and does not alter the Hamiltonian in the central region of a sufficiently long ladder.

\begin{table}[t]
	\centering
	\caption{
		Boundary parameters $(J_e,r)$ used in the calculations of the spin gap at $J=0.4$.
	}
	\label{tab:spin_gap_boundary_parameters}
	\renewcommand{\arraystretch}{1.15}
	\begin{tabular}{c@{\hspace{0.2cm}}c@{\hspace{0.5cm}}c@{\hspace{0.2cm}}c}
		\hline\hline
		\multicolumn{2}{c@{\hspace{0.5cm}}}{$J_{\mathrm H}=0.5$}
		&
		\multicolumn{2}{c@{}}{$J_{\mathrm H}=3.0$}
		\\
		\cline{1-2}\cline{3-4}
		$\delta$ & $(J_e,r)$ & $\delta$ & $(J_e,r)$
		\\
		\hline
		$0$     & $(0,0)$       & $0$     & $(0,0)$       \\
		$0.167$ & $(0,0)$       & $0.083$ & $(0,0)$       \\
		$0.333$ & $(1.6,0.5)$   & $0.167$ & $(0,0)$       \\
		$0.500$ & $(1.6,0.3)$   & $0.250$ & $(0.8,0.5)$   \\
		$0.667$ & $(2.4,0.5)$   & $0.333$ & $(0.8,0.5)$   \\
		$0.750$ & $(3.0,0.7)$   & $0.500$ & $(2.4,0.5)$   \\
		&               & $0.625$ & $(2.6,0.5)$   \\
		\hline\hline
	\end{tabular}
\end{table}

For each parameter set $(J,J_{\mathrm H},\delta)$, we first search for a pair $(J_e,r)$ that removes the boundary-localized excitation and produces a bulk-like $\delta m_i$ profile for every system size included in the finite-size scaling. It is important to use a common pair $(J_e,r)$ throughout a given size sequence; otherwise, a size-dependent boundary condition could contaminate the extrapolation. The boundary parameters listed in Table~\ref{tab:spin_gap_boundary_parameters} were selected according to this criterion.

The acceptable boundary parameters are not unique. In general, we find a finite window of $(J_e,r)$ values for which the boundary-localized contribution is suppressed and the extrapolated bulk gap remains essentially unchanged. This window becomes narrower with increasing hole concentration. At large $\delta$, the smaller bulk spin gap and the stronger sensitivity of the boundary charge and spin environment make it increasingly difficult to find a single pair $(J_e,r)$ that eliminates the boundary-localized excitation over the entire range of system sizes. The boundary profiles must therefore be examined separately for every parameter set and every size used in the extrapolation. Nevertheless, for all parameter sets listed in Table~\ref{tab:spin_gap_boundary_parameters}, such a pair was found, as verified from the $\delta m_i$ profiles at every size used.

Within the resulting admissible window, moderate changes in $J_e$ and $r$, as well as in the cutoff used to terminate the boundary modification, leave the bulk-extrapolated gap essentially unchanged. These checks confirm that the extracted gap represents a bulk energy scale rather than one introduced by a particular choice of boundary parameters.

The terminal-site removal illustrated in Fig.~\ref{fig:spin_boundary}(b) is applied in all finite-size spin-gap calculations. The notation $(J_e,r)=(0,0)$ indicates that no additional exchange tapering is applied; the two terminal sites are nevertheless removed.

We fit the finite-size dependence using
\begin{equation}
	\Delta_s(L)
	=
	\Delta_s(\infty)
	+
	\frac{b}{L^2}
	+
	\frac{a}{L^3},
	\label{eq:spin_gap_scaling}
\end{equation}
where $\Delta_s(\infty)$ is the thermodynamic-limit spin gap. The leading $L^{-2}$ correction is motivated by the momentum quantization of a massive spin excitation whose dispersion has a quadratic expansion near its minimum, while the $L^{-3}$ term allows for the next-order correction. The fits use system sizes for which the boundary-modified regions do not overlap and a well-defined unmodified bulk region remains.

\begin{figure}[bt]
	\centering
	\includegraphics[width=0.7\linewidth]{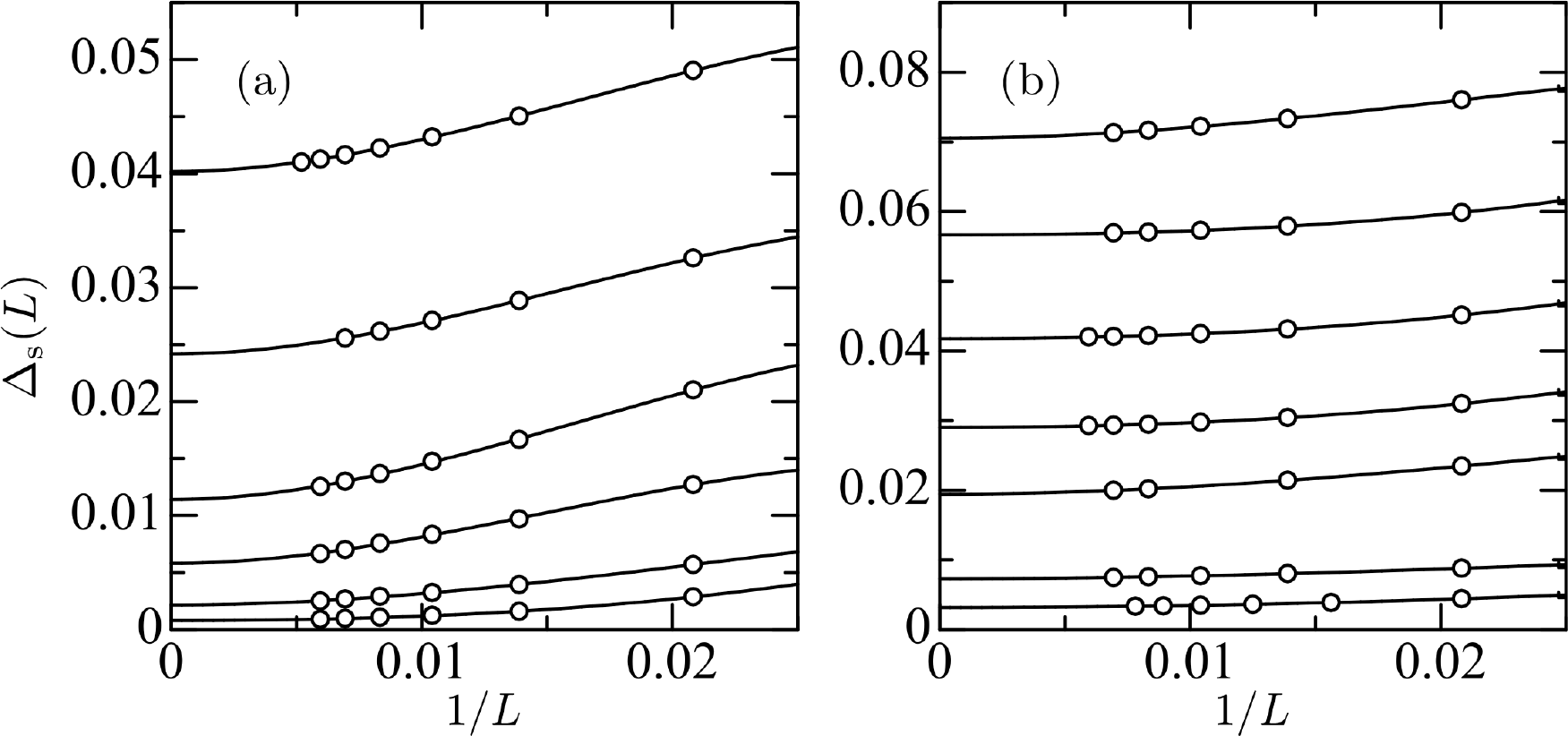}
	\caption{
		Finite-size scaling of the spin gap $\Delta_s(L)$ at $J=0.4$ for
		(a) $J_{\mathrm H}=0.5$ and (b) $J_{\mathrm H}=3.0$.
		The length $L$ denotes the number of rung positions. The solid curves
		are fits to Eq.~(\ref{eq:spin_gap_scaling}), and their intercepts at
		$1/L=0$ give the thermodynamic-limit spin gaps. The boundary
		parameters used for the indicated hole concentrations are listed in
		Table~\ref{tab:spin_gap_boundary_parameters}. Energies are measured
		in units of $t=1$.
	}
	\label{fig:spin_gap_scaling}
\end{figure}

The resulting finite-size extrapolations are shown in Fig.~\ref{fig:spin_gap_scaling}, with the boundary parameters summarized in Table~\ref{tab:spin_gap_boundary_parameters}. For $J_{\mathrm H}=0.5$, the extrapolated spin gap decreases rapidly with increasing doping and becomes very small near the high-doping boundary. For $J_{\mathrm H}=3.0$, the extrapolated spin gap remains clearly finite up to $\delta=0.625$, immediately below the transition at $\delta_c\simeq0.63$. These results confirm that the nonzero spin gaps reported in the main text represent bulk excitation gaps rather than residual boundary-energy scales.

\section{Gapless charge mode and central charge}
\label{sec:central_charge}

To characterize the gapless sector of the doped Haldane regime,
we calculate the von Neumann entanglement entropy using finite
DMRG with open boundary conditions. We employ the boundary
treatment described in Sec.~\ref{sec:spin_gap_boundary}, with
the corresponding boundary parameters $(J_e,r)$.
For a bipartition at the bond between rungs $\ell$ and $\ell+1$,
the entropy is defined as
\begin{equation}
	S(\ell)
	=
	-\mathrm{Tr}\,\rho_\ell\ln\rho_\ell,
	\label{eq:von_Neumann_entropy}
\end{equation}
where $\rho_\ell$ is the reduced density matrix of the subsystem
to the left of the cut.

For a critical one-dimensional system of length $L$ with open
boundaries, the leading conformal scaling form is
\begin{equation}
	S(\ell)
	=
	\frac{c}{6}
	\ln\left[
	\frac{2L}{\pi}
	\sin\left(\frac{\pi\ell}{L}\right)
	\right]
	+s_{\mathrm b},
	\label{eq:central_charge_OBC}
\end{equation}
where $c$ is the central charge and $s_{\mathrm b}$ is a
nonuniversal constant that includes the boundary contribution
\cite{Calabrese2004,Lauchli2008}.
Localized boundary modifications can change the constant and
introduce finite-size corrections, but the asymptotic logarithmic
coefficient is determined by the bulk critical theory.
In a system with both gapless and massive sectors, the use of
Eq.~(\ref{eq:central_charge_OBC}) additionally requires that
the entropy contribution from the massive sectors be approximately
constant over the fitting region.

Open boundaries induce Friedel oscillations in the density and
corresponding oscillations in the entanglement entropy.
For an electron density $n=1-\delta$, the relevant wave vectors are
\begin{equation}
	q_{2k_F}=2k_F=\pi n,
	\qquad
	q_{4k_F}=4k_F=2\pi n
	\equiv-2\pi\delta
	\pmod{2\pi}.
	\label{eq:friedel_wavevector}
\end{equation}
In the present data, the dominant oscillation is the $4k_F$
component for $\delta=1/6$ and $1/3$, with periods of six and
three rungs, respectively. At $\delta=1/2$, the dominant
oscillation is the $2k_F$ component, with wave vector $\pi/2$
and a period of four rungs.

\begin{figure}[t]
	\centering
	\includegraphics[width=0.6\linewidth]{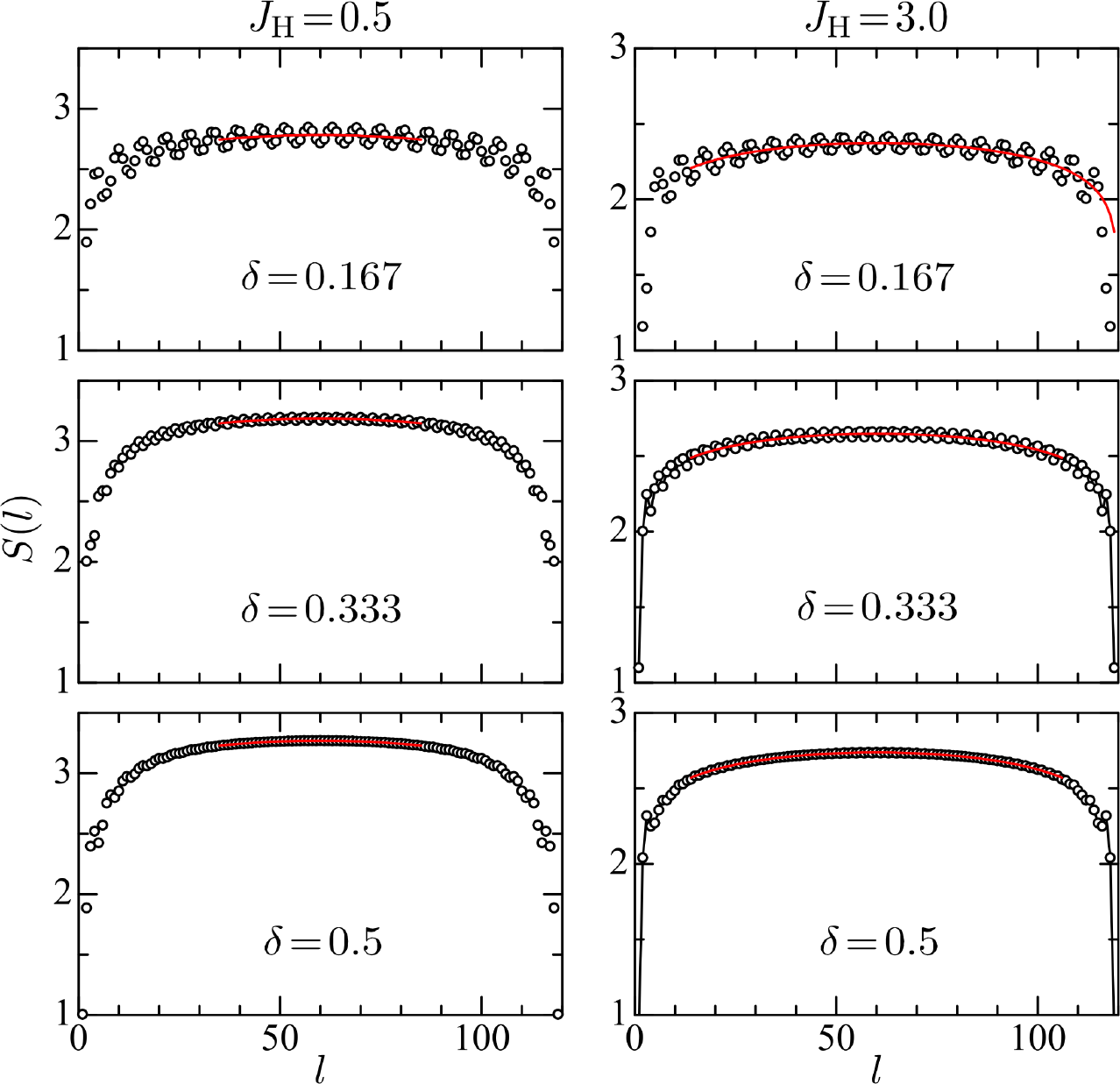}
	\caption{
		Von Neumann entanglement entropy $S(\ell)$ for open
		ladders of length $L=121$ at $J=0.4$.
		The left and right columns show results for
		$J_{\rm H}=0.5$ and $3.0$, respectively, at the
		indicated hole concentrations.
		Symbols show the unaveraged DMRG data.
		The red curves represent the conformal profiles obtained
		from the moving-average fits in
		Eq.~(\ref{eq:central_charge_fit}).
		Fit ranges, specified by the window center $\ell_c$,
		are given in Table~\ref{tab:central_charge_SM}.
	}
	\label{fig:central_charge_SM}
\end{figure}

Before fitting, we average the entropy over one period of the
dominant oscillation:
\begin{equation}
	\overline S_w(\ell)
	=
	\frac{1}{w}
	\sum_{j=0}^{w-1}S(\ell+j).
	\label{eq:entropy_moving_average}
\end{equation}
We define the logarithmic conformal distance as
\begin{equation}
	x(\ell)
	=
	\ln\left[
	\frac{2L}{\pi}
	\sin\left(\frac{\pi\ell}{L}\right)
	\right]
	\label{eq:conformal_distance}
\end{equation}
and average it over the same window:
\begin{equation}
	\overline x_w(\ell)
	=
	\frac{1}{w}
	\sum_{j=0}^{w-1}x(\ell+j).
	\label{eq:conformal_distance_average}
\end{equation}
The fitted central-charge estimate $c_{\mathrm{fit}}$ is obtained
from the linear fit
\begin{equation}
	\overline S_w(\ell)
	\simeq
	\frac{c_{\mathrm{fit}}}{6}\,
	\overline x_w(\ell)
	+s_{\mathrm b}.
	\label{eq:central_charge_fit}
\end{equation}
We use $w=6$, $3$, and $4$ for $\delta=1/6$, $1/3$, and
$1/2$, respectively. Averaging both quantities preserves the
linear relation between them when the conformal form applies.

Each averaged point is assigned to the center of its window,
$\ell_c=\ell+(w-1)/2$.
The quoted fit ranges refer to $\ell_c$, rather than to the
leftmost cut $\ell$. The cuts contributing to the fit can
therefore extend by up to $(w-1)/2$ beyond either end of
the quoted interval.

Figure~\ref{fig:central_charge_SM} shows the entropy for three
representative hole concentrations at each Hund coupling.
For $J_{\rm H}=3.0$, we use the window-center range
$14\leq\ell_c\leq106$. The cuts contributing to these averages
lie outside the region of enhanced boundary exchange.
For $J_{\rm H}=0.5$, we use the narrower central range
$35\leq\ell_c\leq85$ to reduce boundary effects and residual
finite-distance contributions from the massive spin sector.

The narrower range at weak Hund coupling is motivated by the
small bulk spin gap and the spatial crossover observed in the
local slope of the moving-averaged entropy as a function of
the averaged conformal distance. This crossover occurs over
several tens of rungs and guides the choice of the central
fitting region. It is not an independent determination of
the spin correlation length, and restricting the fit to this
region does not guarantee that all finite-size corrections
have been removed.

The selected-window estimates are summarized in
Table~\ref{tab:central_charge_SM}.
All six values lie between $0.93$ and $1.08$ and are
consistent with $c=1$. Their sensitivity to the fitting
interval, however, differs between the two Hund couplings.

For $J_{\rm H}=3.0$, varying the fitting interval over the
ranges examined gives estimates between approximately
$0.93$ and $1.07$ across the three dopings. The persistence
of values close to unity under these variations provides
robust finite-size evidence for $c\simeq1$.
For $J_{\rm H}=0.5$, the estimates at $\delta=1/3$ and
$1/2$ depend appreciably on the fitting interval.
The central-window values in Table~\ref{tab:central_charge_SM}
are compatible with $c=1$, but the window dependence limits
the precision with which the asymptotic central charge can
be inferred from the available system size.

\begin{table}[t]
	\centering
	\caption{
		Finite-size central-charge estimates from
		Eq.~(\ref{eq:central_charge_fit}) at $J=0.4$.
		The selected fit ranges are
		$35\leq\ell_c\leq85$ for $J_{\rm H}=0.5$ and
		$14\leq\ell_c\leq106$ for $J_{\rm H}=3.0$.
		Here, $\ell_c$ denotes the center of each averaging
		window. The values are fit estimates for these
		intervals, rather than thermodynamic-limit
		extrapolations.
	}
	\label{tab:central_charge_SM}
	\renewcommand{\arraystretch}{1.1}
	\begin{tabular}{ccc}
		\hline\hline
		$\delta$
		& $c_{\mathrm{fit}}\;(J_{\rm H}=0.5)$
		& $c_{\mathrm{fit}}\;(J_{\rm H}=3.0)$ \\
		\hline
		$1/6$ & $0.97$ & $0.97$ \\
		$1/3$ & $1.03$ & $0.93$ \\
		$1/2$ & $1.08$ & $0.97$ \\
		\hline\hline
	\end{tabular}
\end{table}

Together with the independently established finite bulk
spin gap, these results support a $C1S0$ description with
one gapless charge mode and a gapped spin sector.
The evidence from entropy scaling is more robust for
$J_{\rm H}=3.0$, whereas the weak-Hund results retain
larger finite-size uncertainties.
This interpretation is consistent with the bosonization
picture in Sec.~\ref{sec:SM_bosonization}.
On the strong-Hund side, a single gapless charge mode is
also compatible with the formation of mobile rung-triplet
pairs discussed in Sec.~\ref{sec:dilute_analysis}.

\section{Entanglement-spectrum analysis}
\label{sec:entanglement_spectrum}

For a bipartition of the infinite ladder across an inter-rung bond, the Schmidt decomposition of the ground state is written as
\begin{equation}
	|\Psi\rangle
	=
	\sum_{\alpha}
	\lambda_{\alpha}
	|\alpha\rangle_{\mathrm{L}}
	|\alpha\rangle_{\mathrm{R}},
	\qquad
	\sum_{\alpha}\lambda_{\alpha}^{2}=1.
	\label{eq:schmidt}
\end{equation}
The entanglement spectrum (ES) is defined from the Schmidt coefficients as
\begin{equation}
	\mathcal{E}_{\alpha}
	=
	-\ln\lambda_{\alpha}^{2}.
	\label{eq:ES}
\end{equation}
Because the calculations conserve the particle number and the total spin projection, each Schmidt state is labeled by the corresponding bond quantum numbers $Q_N$ and $Q_{2S^z}$. Here, $Q_{2S^z}$ is the integer-valued spin quantum number corresponding to twice the spin projection. The overall offsets of these bond quantum numbers depend on the MPS convention; only sector differences, degeneracies, and their evolution between cuts have physical significance.

For the representative calculations discussed in this section, the MPS unit cell contains eight rungs, or equivalently 16 MPS sites. We analyze the eight even MPS bonds, $b=0,2,\ldots,14$, which correspond to the inequivalent inter-rung cuts. More generally, the number of inequivalent inter-rung cuts equals the number of rungs in the iDMRG unit cell. The absolute charge label associated with the largest Schmidt coefficient is not generally the same at different cuts. In particular, even when the dominant sector is labeled $Q_N=0$ at one cut, the corresponding sector at another cut can have a different $Q_N$ because the conserved bond charge changes as the cut is moved through the unit cell. We therefore do not impose the same numerical value of $Q_N$ at every cut.

For each inter-rung cut $b$, we identify the dominant particle-number sector $Q_N^{(b)}$ as the sector containing the largest Schmidt coefficient, or equivalently the lowest entanglement level. We then collect all $Q_{2S^z}$ subsectors within this fixed particle-number sector and order their levels according to increasing $\mathcal{E}_{\alpha}$. Because particle-number conservation decomposes the reduced density matrix into independent charge blocks, whereas the Haldane entanglement structure resides in the spin sector, the even-fold organization is examined within a fixed charge block rather than by mixing levels belonging to different particle-number sectors.

The $k$th level at cut $b$ is denoted by $\mathcal{E}_{k}^{(b)}$. The cut-averaged level is defined as
\begin{equation}
	\overline{\mathcal{E}}_{k}
	=
	\frac{1}{N_{\mathrm{cut}}}
	\sum_b
	\mathcal{E}_{k}^{(b)},
	\qquad
	N_{\mathrm{cut}}=8.
	\label{eq:ES_average}
\end{equation}
At sufficiently large bond dimension, the organization and spin-sector assignment of the low-lying levels are consistent among the eight cuts, making this rank-by-rank correspondence unambiguous.

Table~\ref{tab:ES_levels} gives a representative example at $J=0.4$, $J_{\mathrm H}=3.0$, $n=0.625$ ($\delta=0.375$), and $\chi=6000$. For this parameter set, the dominant charge sectors at the eight inter-rung cuts are
\begin{equation}
	Q_N^{(b)}
	=
	(0,0,2,4,4,6,8,8)
	\label{eq:dominant_sectors_JH3}
\end{equation}
for $b=(0,2,4,6,8,10,12,14)$, respectively. To compare the different cuts, we introduce the relative charge label
\begin{equation}
	\widetilde{Q}_N
	=
	Q_N-Q_N^{(b)}
	\label{eq:relative_charge_ES}
\end{equation}
at each cut. Thus, $\widetilde{Q}_N=0$ denotes the dominant particle-number sector independently of its absolute bond-charge label.

\begin{table}[t]
	\centering
	\caption{
		Lowest 20 cut-averaged entanglement levels at $J=0.4$, $J_{\mathrm H}=3.0$, $n=0.625$ ($\delta=0.375$), and $\chi=6000$. The charge quantum number $\widetilde{Q}_N$ is measured relative to the dominant particle-number sector at each cut.
	}
	\label{tab:ES_levels}
	\begin{tabular}{
			c@{\hspace{1.2cm}}
			c@{\hspace{1.2cm}}
			c@{\hspace{1.2cm}}
			c
		}
		\hline\hline
		$k$ & $\overline{\mathcal{E}}_k$
		& $\widetilde{Q}_N$ & $Q_{2S^z}$ \\
		\hline
		1  & 1.578840 & 0 &  0 \\
		2  & 1.578840 & 0 &  2 \\
		3  & 3.869461 & 0 &  0 \\
		4  & 3.869461 & 0 &  2 \\
		5  & 5.140462 & 0 & $-2$ \\
		6  & 5.140463 & 0 &  4 \\
		7  & 5.140487 & 0 &  0 \\
		8  & 5.140487 & 0 &  2 \\
		9  & 6.283653 & 0 &  0 \\
		10 & 6.283653 & 0 &  2 \\
		11 & 6.563358 & 0 &  0 \\
		12 & 6.563358 & 0 &  2 \\
		13 & 7.135085 & 0 &  0 \\
		14 & 7.135085 & 0 &  2 \\
		15 & 7.488552 & 0 &  2 \\
		16 & 7.488554 & 0 &  0 \\
		17 & 7.637180 & 0 & $-2$ \\
		18 & 7.637183 & 0 &  4 \\
		19 & 7.637232 & 0 &  0 \\
		20 & 7.637234 & 0 &  2 \\
		\hline\hline
	\end{tabular}
\end{table}

The lowest two levels in Table~\ref{tab:ES_levels}, for example, belong to different spin sectors but are degenerate within the numerical precision. The higher levels similarly form nearly degenerate doublets or quartets. This systematic even-fold organization within a fixed charge sector is the characteristic entanglement structure inherited from the Haldane spin sector.
We interpret this even-fold organization together with the
string order, bulk spin gap, and boundary-spin signatures.
The ES within a selected charge sector is not used as a
standalone topological invariant, and a symmetry-fractionalization
invariant is not evaluated independently in this work.

We further examine the convergence of the lowest doublet with the MPS bond dimension. For each cut, we select the dominant charge sector at the largest bond dimension, $\chi=6000$, and track the same absolute charge sector toward lower $\chi$. Fixing the sector in this manner is important because two nearby charge sectors can occasionally exchange their ordering at smaller bond dimensions. Selecting the dominant sector independently at every $\chi$ could therefore result in the comparison of different sequences of entanglement levels.

Within the fixed sector $Q_N^{(b)}$, we define the splitting between the two lowest levels at cut $b$ as
\begin{equation}
	\Delta_{\mathrm{ES}}^{(b)}(\chi)
	=
	\mathcal{E}_{2}^{(b)}(\chi)
	-
	\mathcal{E}_{1}^{(b)}(\chi),
	\label{eq:Delta_ES_cut}
\end{equation}
and its cut-averaged value as
\begin{equation}
	\Delta_{\mathrm{ES}}(\chi)
	=
	\frac{1}{N_{\mathrm{cut}}}
	\sum_b
	\Delta_{\mathrm{ES}}^{(b)}(\chi).
	\label{eq:Delta_ES}
\end{equation}
The resulting bond-dimension dependence is shown in Table~\ref{tab:ES_chi}. We consider the two representative Hund couplings $J_{\mathrm H}=0.5$ and $J_{\mathrm H}=3.0$ at fixed $J=0.4$ and $n=0.625$. At $\chi=6000$, the dominant charge sectors for $J_{\mathrm H}=0.5$ are
\begin{equation}
	Q_N^{(b)}
	=
	(0,2,2,4,4,6,8,8),
	\label{eq:dominant_sectors_JH05}
\end{equation}
whereas those for $J_{\mathrm H}=3.0$ are given in Eq.~(\ref{eq:dominant_sectors_JH3}).

\begin{table}[t]
	\centering
	\caption{
		Bond-dimension dependence of the cut-averaged splitting $\Delta_{\mathrm{ES}}$ between the two lowest entanglement levels at $J=0.4$ and $n=0.625$ ($\delta=0.375$). The average is taken over the eight inequivalent inter-rung cuts.
	}
	\label{tab:ES_chi}
	\begin{tabular}{
			c@{\hspace{1.2cm}}
			c@{\hspace{1.2cm}}
			c
		}
		\hline\hline
		$\chi$ &
		$\Delta_{\mathrm{ES}}\;(J_{\mathrm H}=0.5)$ &
		$\Delta_{\mathrm{ES}}\;(J_{\mathrm H}=3.0)$ \\
		\hline
		500  & $4.862\times10^{-3}$ & $1.083\times10^{-4}$ \\
		1000 & $1.140\times10^{-4}$ & $6.050\times10^{-5}$ \\
		1500 & $7.813\times10^{-5}$ & $6.750\times10^{-6}$ \\
		2000 & $1.379\times10^{-4}$ & $1.236\times10^{-4}$ \\
		2500 & $3.875\times10^{-6}$ & $4.250\times10^{-6}$ \\
		3000 & $3.375\times10^{-6}$ & $\lesssim10^{-6}$ \\
		3500 & $1.950\times10^{-5}$ & $6.250\times10^{-6}$ \\
		4000 & $1.413\times10^{-5}$ & $\lesssim10^{-6}$ \\
		4500 & $2.000\times10^{-6}$ & $\lesssim10^{-6}$ \\
		5000 & $2.375\times10^{-6}$ & $\lesssim10^{-6}$ \\
		5500 & $2.625\times10^{-5}$ & $\lesssim10^{-6}$ \\
		6000 & $9.375\times10^{-6}$ & $\lesssim10^{-6}$ \\
		\hline\hline
	\end{tabular}
\end{table}

Entries denoted by $\lesssim10^{-6}$ in Table~\ref{tab:ES_chi} are below the resolution set by the six decimal places retained for the entanglement levels in the numerical output. The convergence of $\Delta_{\mathrm{ES}}$ is not strictly monotonic, reflecting residual finite-$\chi$ and optimization effects at the level of these very small splittings. Nevertheless, at the largest bond dimensions the splitting is reduced to approximately $10^{-5}$ for $J_{\mathrm H}=0.5$ and to the numerical resolution of the extracted spectrum for $J_{\mathrm H}=3.0$. These values are several orders of magnitude smaller than the separations between distinct low-lying entanglement multiplets. We therefore conclude that the even-fold low-energy organization of the ES is well converged at the bond dimensions used in the main analysis.

\section{Distance dependence and discarded-weight extrapolation of the string-order parameter}
\label{sec:string_order}

To examine the persistence of the Haldane-type order upon hole doping, we calculate the string-order correlation function
\begin{equation}
	O_{\mathrm{str}}(r)
	=
	-
	\left\langle
	\widetilde{S}^{z}_{i}
	\exp\left(
	i\pi
	\sum_{j=i+1}^{i+r-1}
	\widetilde{S}^{z}_{j}
	\right)
	\widetilde{S}^{z}_{i+r}
	\right\rangle,
	\label{eq:string_correlation}
\end{equation}
where
\begin{equation}
	\widetilde{S}^{z}_{i}
	=
	S^{z}_{i,1}+S^{z}_{i,2}
\end{equation}
is the total spin projection on rung $i$. In the iDMRG calculations, Eq.~(\ref{eq:string_correlation}) is averaged over the inequivalent choices of the initial rung $i$ within the MPS unit cell. The string-order parameter is defined by the long-distance limit
\begin{equation}
	O_{\mathrm{str}}
	=
	\lim_{r\rightarrow\infty}
	O_{\mathrm{str}}(r).
	\label{eq:string_order}
\end{equation}

\begin{figure}[t]
	\centering
	\includegraphics[width=0.9\linewidth]{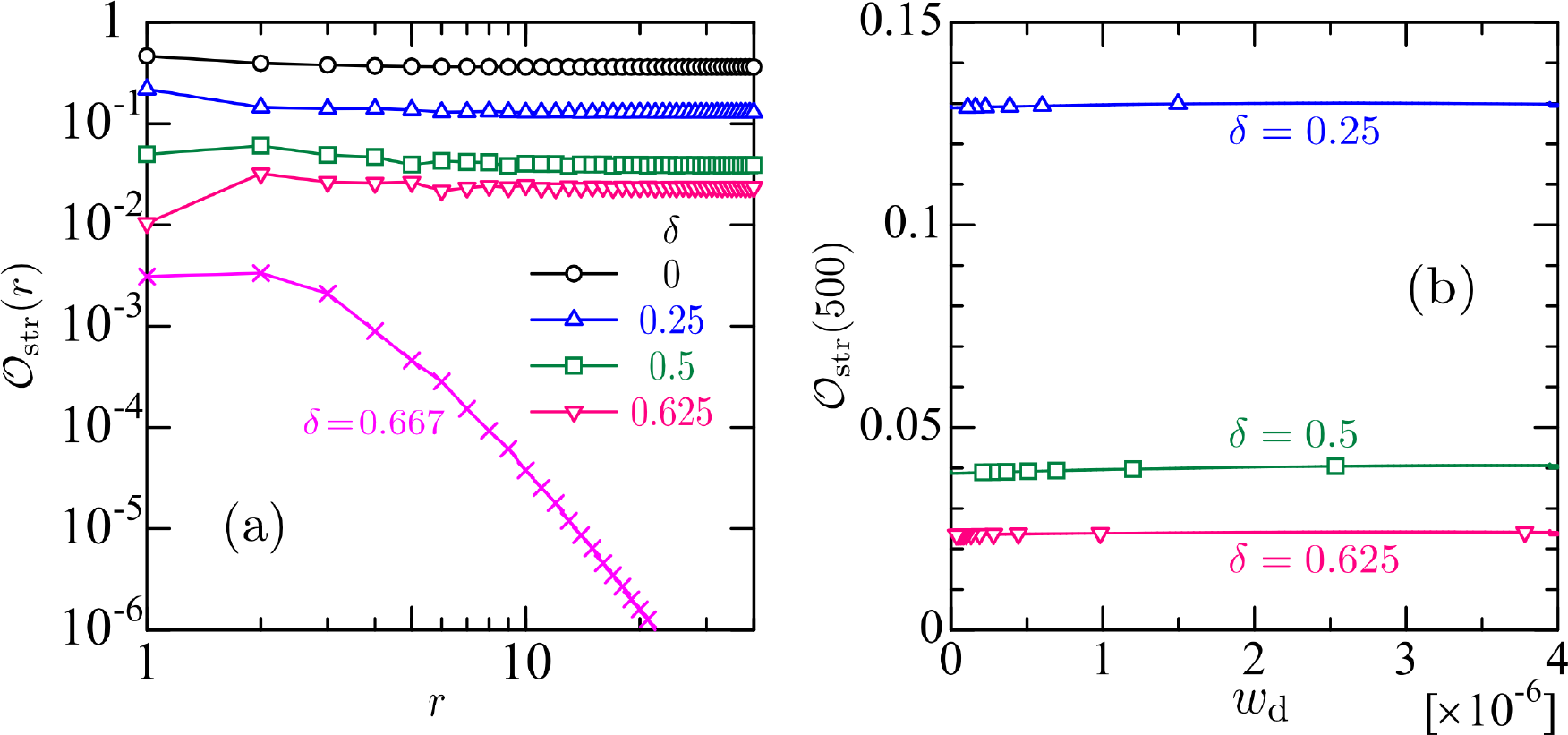}
	\caption{
		String-order correlation at $J=0.4$ and $J_{\mathrm H}=3.0$.
		(a) Distance dependence of $O_{\mathrm{str}}(r)$ for the indicated hole concentrations at $\chi=6000$.
		(b) Long-distance value $O_{\mathrm{str}}(r=500)$ as a function of the discarded weight $w_{\mathrm d}$. The solid lines are linear fits to the data at small $w_{\mathrm d}$.
	}
	\label{fig:string_order_SM}
\end{figure}

Figure~\ref{fig:string_order_SM}(a) shows the distance dependence of
$O_{\mathrm{str}}(r)$ for $J=0.4$ and $J_{\mathrm H}=3.0$. For
$\delta\leq0.625$, the correlation function approaches a nonzero
plateau at long distances, although its magnitude decreases with
increasing hole concentration. By contrast, at $\delta=0.667$,
$O_{\mathrm{str}}(r)$ decays rapidly toward zero. The contrasting
behaviors at $\delta=0.625$ and $0.667$ therefore locate the
disappearance of the string order between these two dopings,
consistent with the phase boundary obtained from the other
diagnostics.

We next examine whether the finite long-distance values are stable
against the finite MPS truncation. As demonstrated by the plateaus in
Fig.~\ref{fig:string_order_SM}(a), $r=500$ is sufficiently large to
represent the long-distance regime for the dopings considered in the
extrapolation. Figure~\ref{fig:string_order_SM}(b) shows
$O_{\mathrm{str}}(r=500)$ as a function of the discarded weight
$w_{\mathrm d}$ for representative dopings, including
$\delta=0.625$ close to the phase boundary. Here, $w_{\mathrm d}$
denotes the maximum discarded weight in the final converged iDMRG
sweep over the bonds of the MPS unit cell.

The data in the small-$w_{\mathrm d}$ regime exhibit an approximately
linear dependence and are fitted using
\begin{equation}
	O_{\mathrm{str}}(500,w_{\mathrm d})
	=
	O_{\mathrm{str}}^{(0)}(500)
	+
	a w_{\mathrm d},
	\label{eq:string_extrapolation}
\end{equation}
where $O_{\mathrm{str}}^{(0)}(500)$ denotes the estimate in the
zero-discarded-weight limit. The extrapolated values remain finite
for all three dopings shown in Fig.~\ref{fig:string_order_SM}(b),
including $\delta=0.625$. Together with the long-distance saturation
shown in Fig.~\ref{fig:string_order_SM}(a), this demonstrates that the
finite string order below the transition is stable against both the
finite separation and the finite MPS truncation.

\section{Bosonization argument for the doped Haldane regime}
\label{sec:SM_bosonization}

We summarize the field-theoretical interpretation of the doped Haldane regime based on the weak-coupling analysis of Fujimoto and Kawakami~\cite{Fujimoto1995}. Our purpose is not to derive a new continuum theory or to determine the phase boundary analytically, but to explain how a gapless charge mode can coexist with a gapped spin sector retaining Haldane-type string order.

Fujimoto and Kawakami considered two itinerant Hubbard chains coupled by a ferromagnetic on-site Hund interaction, without single-particle hopping between the chains. Our projected two-leg $t$--$J$ ladder is a strong-coupling counterpart of this model. Although the weak-coupling analysis is not quantitatively controlled in the parameter regime studied here, it provides a useful description of the separation between the charge and spin sectors.

Let $a=1,2$ denote the chain index and $\alpha=\uparrow,\downarrow$ the spin index. The smooth part of the density of each species can be expressed as
\begin{equation}
	\rho_{a\alpha}(x)
	=
	\rho_0
	-
	\frac{1}{\pi}
	\partial_x\phi_{a\alpha}(x)
	+\cdots,
	\label{eq:SM_density}
\end{equation}
where the ellipsis denotes oscillatory contributions. We introduce charge and spin fields for each chain,
\begin{equation}
	\phi_{a\rho}
	=
	\frac{
		\phi_{a\uparrow}+\phi_{a\downarrow}
	}{
		\sqrt{2}
	},
	\qquad
	\phi_{a\sigma}
	=
	\frac{
		\phi_{a\uparrow}-\phi_{a\downarrow}
	}{
		\sqrt{2}
	},
	\label{eq:SM_charge_spin}
\end{equation}
and their symmetric and antisymmetric combinations,
\begin{equation}
	\phi_{\rho\pm}
	=
	\frac{
		\phi_{1\rho}\pm\phi_{2\rho}
	}{
		\sqrt{2}
	},
	\qquad
	\phi_{\sigma\pm}
	=
	\frac{
		\phi_{1\sigma}\pm\phi_{2\sigma}
	}{
		\sqrt{2}
	}.
	\label{eq:SM_pm_fields}
\end{equation}
The corresponding dual fields are denoted by $\theta_{\rho\pm}$ and $\theta_{\sigma\pm}$. The four modes are therefore the total charge mode $(\rho+)$, the relative charge mode $(\rho-)$, the symmetric spin mode $(\sigma+)$, and the antisymmetric spin mode $(\sigma-)$.

A chemical potential couples to the total density according to
\begin{equation}
	H_{\mu}
	=
	-\mu
	\int dx\,
	\rho_{\mathrm{tot}}(x),
	\qquad
	\rho_{\mathrm{tot}}(x)
	=
	\sum_{a,\alpha}
	\rho_{a\alpha}(x).
	\label{eq:SM_chemical_potential}
\end{equation}
Using Eq.~(\ref{eq:SM_density}), its smooth contribution is
\begin{equation}
	\rho_{\mathrm{tot}}(x)
	=
	4\rho_0
	-
	\frac{2}{\pi}
	\partial_x\phi_{\rho+}(x)
	+\cdots,
	\label{eq:SM_total_density}
\end{equation}
and hence
\begin{equation}
	H_{\mu}
	=
	\frac{2\mu}{\pi}
	\int dx\,
	\partial_x\phi_{\rho+}(x)
	+
	\mathrm{const.}
	\label{eq:SM_mu_rhoplus}
\end{equation}
The chemical potential therefore changes the background value of the total charge field but does not directly couple to the relative charge or spin fields.

At half filling, a commensurate Umklapp interaction can gap the total charge sector. Schematically, this contribution has the form
\begin{equation}
	H_{\mathrm{u}}
	\sim
	g_{\mathrm{u}}
	\int dx\,
	\cos\left[
	\beta_{\rho+}\phi_{\rho+}(x)
	-
	Q_{\delta}x
	\right],
	\label{eq:SM_umklapp}
\end{equation}
where $Q_{\delta}=0$ at the commensurate filling and becomes nonzero upon doping. Away from half filling, the spatially oscillating Umklapp term does not pin $\phi_{\rho+}$ at long distances. The remaining total charge sector is then described by a Gaussian Hamiltonian,
\begin{equation}
	H_{\rho+}
	=
	\frac{v_{\rho+}}{2\pi}
	\int dx\,
	\left[
	K_{\rho+}
	\left(
	\partial_x\theta_{\rho+}
	\right)^2
	+
	\frac{1}{K_{\rho+}}
	\left(
	\partial_x\phi_{\rho+}
	\right)^2
	\right],
	\label{eq:SM_total_charge_gaussian}
\end{equation}
which represents one gapless charge mode with central charge $c=1$.

The absence of a gap in the total charge sector does not require the other three modes to become gapless. The interactions generated by the ferromagnetic Hund coupling contain nonoscillating cosine operators involving the relative charge and spin fields. In the Haldane-like strong-coupling branch identified by Fujimoto and Kawakami, these interactions generate masses in the relative charge mode, the symmetric spin mode, and the antisymmetric spin mode. In terms of the fields defined above, the corresponding locking pattern can be represented schematically as
\begin{equation}
	\phi_{\rho-},
	\qquad
	\phi_{\sigma+},
	\qquad
	\theta_{\sigma-}
	\quad
	\text{pinned},
	\label{eq:SM_Haldane_pinning}
\end{equation}
while $\phi_{\rho+}$ remains unpinned. The precise pinned values and the coefficients of the cosine operators depend on the bosonization convention. This fixed-point structure contains one gapless charge mode and no gapless spin mode and is therefore a $C1S0$ state.

The relation between this locking pattern and Haldane order can be seen from the nonlocal string correlator introduced by den Nijs and Rommelse~\cite{denNijs1989}, whose connection to hidden symmetry breaking is made explicit by the Kennedy--Tasaki transformation~\cite{Kennedy1992}. In the bosonized description, the long-distance string order is associated with an operator in the symmetric spin sector. Schematically,
\begin{equation}
	\mathcal{O}_{\mathrm{str}}
	\sim
	\left|
	\left\langle
	\cos\left(
	\beta_s\phi_{\sigma+}
	\right)
	\right\rangle
	\right|^2,
	\label{eq:SM_string_bosonized}
\end{equation}
where $\beta_s$ and the overall proportionality factor depend on the normalization convention. The squared expectation value reflects the factorization of the corresponding two-point function in the long-distance limit. The essential implication of Eq.~(\ref{eq:SM_string_bosonized}) is that the string order can remain nonzero as long as the symmetric spin field remains pinned.

This field-theoretical picture explains how carrier doping can metallize the system without immediately destroying its Haldane spin structure. Doping removes the commensurate locking of the total charge field, whereas the interactions that gap the relative charge and spin sectors can remain effective. The resulting state has a gapless total charge mode coexisting with a finite spin gap, finite string order, and the characteristic even-fold organization of the entanglement spectrum.

This interpretation is consistent with our numerical results.
The central-charge estimates in Sec.~\ref{sec:central_charge}
are consistent with $c\simeq1$. Together with the finite bulk spin gap established in Sec.~\ref{sec:spin_gap_boundary}, this identifies the doped state as a $C1S0$ metal. The finite string order reported in Sec.~\ref{sec:string_order} and the entanglement-spectrum structure discussed in Sec.~\ref{sec:entanglement_spectrum} further show that the gapped spin sector retains the characteristic Haldane structure. For example, at $J=0.4$ and $J_{\mathrm H}=3.0$, the string order remains finite up to $\delta=0.625$, immediately below the numerically determined phase boundary.

We emphasize that doping can also renormalize the couplings and scaling dimensions of the locking operators. The above argument therefore does not imply that the spin and relative-charge gaps must survive at arbitrary doping, nor does it determine the critical hole concentration. At the boundary of the doped Haldane regime, the numerical results cease to support the pinned Haldane spin-sector description, as indicated by the combined evolution of the string order, spin gap, and entanglement spectrum. Depending on the microscopic parameters, this breakdown need not correspond to a simple continuous depinning transition.

The weak-coupling bosonization argument should thus be regarded as a qualitative explanation of the compatibility between metallic charge transport and a gapped Haldane-type spin sector. The DMRG and iDMRG calculations provide the nonperturbative evidence that this structure persists in the strongly correlated $t$--$J$ ladder and determine its range of stability. The persistence of boundary-localized spin excitations in the doped regime is discussed separately in Sec.~\ref{sec:spin_gap_boundary}. More generally, gapless phases carrying symmetry-protected topological structure have been discussed in a variety of one-dimensional settings~\cite{Keselman2015,Scaffidi2017,Keselman2018}.

\section{Local rung bookkeeping and dilute-defect configurations}
\label{sec:local_rung}

Here we provide further details of the local rung mechanism discussed in the main text. In the projected $t$--$J$ model, each site can be either empty or singly occupied. We denote by $P_m$ the probability that a rung contains $m=0,1,$ or $2$ electrons. These probabilities satisfy
\begin{equation}
	P_0+P_1+P_2=1,
	\qquad
	P_1+2P_2=2n,
	\label{eq:SM_rung_constraints}
\end{equation}
where $n=1-\delta$ is the electron density per site. It follows that
\begin{equation}
	P_0=1-2n+P_2,
	\qquad
	P_1=2(n-P_2).
	\label{eq:SM_rung_prob}
\end{equation}

For statistically independent occupations of the two sites on a rung, the corresponding probabilities are
\begin{equation}
	P_2^{(0)}=n^2,
	\qquad
	P_0^{(0)}=(1-n)^2=\delta^2,
	\qquad
	P_1^{(0)}=2n(1-n)=2n\delta.
	\label{eq:SM_uncorrelated_prob}
\end{equation}
Defining the deviations from these uncorrelated values by
\begin{equation}
	\Delta P_m
	=
	P_m-P_m^{(0)},
\end{equation}
Eq.~(\ref{eq:SM_rung_prob}) gives the exact relations
\begin{equation}
	\Delta P_0=\Delta P_2,
	\qquad
	\Delta P_1=-2\Delta P_2.
	\label{eq:SM_rung_sumrule}
\end{equation}
Thus, an enhancement of the probabilities of empty and two-electron rungs relative to an uncorrelated distribution is accompanied by a suppression twice as large in the probability of singly occupied spin-$1/2$ rungs.

Here, $P_2$ denotes the total probability of a two-electron rung and does not by itself distinguish between its singlet and triplet components. In the regime with a sufficiently strong ferromagnetic Hund coupling, the triplet component is dominant, so that most $P_2$ rungs act as local integer-spin units. By contrast, a $P_1$ rung necessarily carries a half-integer spin.

The enhancement of $P_0$ can equivalently be expressed through the normalized same-rung hole correlation. Defining
\begin{equation}
	h_{i,a}=1-n_{i,a},
\end{equation}
we have
\begin{equation}
	g_{hh}(0)
	=
	\frac{
		\left\langle
		h_{i,1}h_{i,2}
		\right\rangle
	}{
		\left\langle h_{i,1}\right\rangle
		\left\langle h_{i,2}\right\rangle
	}
	=
	\frac{P_0}{\delta^2},
	\label{eq:SM_same_rung_hole}
\end{equation}
where translational and leg-exchange symmetry give
$\langle h_{i,a}\rangle=\delta$.

The normalized defect fraction introduced in the main text is defined as
\begin{equation}
	f_{\mathrm{def}}
	=
	\frac{P_1}{P_1^{(0)}}.
	\label{eq:SM_defect_fraction_definition}
\end{equation}
Using Eqs.~(\ref{eq:SM_rung_sumrule}) and
(\ref{eq:SM_same_rung_hole}), we obtain
\begin{equation}
	\Delta P_2
	=
	\Delta P_0
	=
	\delta^2
	\left[
	g_{hh}(0)-1
	\right]
\end{equation}
and
\begin{equation}
	P_1
	=
	P_1^{(0)}-2\Delta P_2.
\end{equation}
Since $P_1^{(0)}=2n\delta$, the normalized defect fraction satisfies the exact relation
\begin{equation}
	f_{\mathrm{def}}
	=
	1-\frac{\Delta P_2}{n\delta}
	=
	1-
	\left[
	g_{hh}(0)-1
	\right]
	\frac{\delta}{n}.
	\label{eq:SM_defect_fraction}
\end{equation}
Thus, the enhancement of the same-rung hole correlation directly quantifies the suppression of the $P_1$ defect-rung probability relative to its uncorrelated value.

This bookkeeping shows why a relatively small absolute density of empty rungs can have a substantial effect on the spin sector. For example, at $\delta=0.05$ and $J_{\mathrm H}=3.0$, we obtain
\begin{equation}
	P_0\simeq0.01985,
	\qquad
	P_0^{(0)}=\delta^2=0.0025,
\end{equation}
and hence
\begin{equation}
	g_{hh}(0)
	=
	\frac{P_0}{\delta^2}
	\simeq7.94.
\end{equation}
The corresponding excess empty-rung and two-electron-rung probabilities are
\begin{equation}
	\Delta P_0
	=
	\Delta P_2
	\simeq0.01735.
\end{equation}
Equation~(\ref{eq:SM_rung_sumrule}) then gives
\begin{equation}
	\Delta P_1
	\simeq-0.03470.
\end{equation}
Since the uncorrelated value is
\begin{equation}
	P_1^{(0)}
	=
	2(0.95)(0.05)
	=
	0.0950,
\end{equation}
the actual probability is approximately
\begin{equation}
	P_1
	\simeq
	0.0603
	\simeq
	0.635P_1^{(0)}.
\end{equation}
Equivalently, Eq.~(\ref{eq:SM_defect_fraction}) gives
\begin{equation}
	f_{\mathrm{def}}
	=
	1-
	(7.94-1)
	\frac{0.05}{0.95}
	\simeq
	0.635,
\end{equation}
in agreement with the ratio obtained directly from the rung probabilities. Thus, although the absolute increase in $P_0$ is only about $0.017$, the probability of singly occupied spin-$1/2$ defect rungs is reduced to approximately $63\%$ of its uncorrelated value.

This redistribution is particularly relevant to the string correlation. An empty $P_0$ rung has $\widetilde{S}_i^z=0$ and therefore contributes the trivial phase factor to the interior of the string operator. A singly occupied $P_1$ rung instead carries a half-integer spin and produces a nontrivial phase factor. Two-electron $P_2$ rungs carry integer total spin, with their triplet component strongly favored by the ferromagnetic Hund coupling. The reduction of $P_1$ rungs produced by the exact relation
$\Delta P_1=-2\Delta P_0$ therefore provides the microscopic basis of the local defect-avoidance mechanism discussed in the main text.

The configurations illustrated in Fig.~1(b) of the main text provide a schematic real-space interpretation of this dilute-hole regime. A local
$P_2$--$P_0$--$P_2$ configuration contains a spinless rung between two spin-carrying two-electron rungs. Upon removing the empty rung in a squeezed-spin-space representation, the surrounding integer-spin backbone remains locally connected. Similarly, configurations containing neighboring $P_1$ rungs, such as
$P_2$--$P_1$--$P_1$--$P_2$, avoid two widely separated isolated half-integer-spin defects and allow the two neighboring defects to contribute collectively as an integer-spin unit. These configurations should be regarded as schematic representations of the short-range correlations observed numerically, rather than as uniquely selected microscopic bound states.

At larger doping, the dilute local-correlation picture alone ceases to provide a sufficient explanation. As shown in Fig.~4(b) of the main text, both the hole and defect-position correlations approach their uncorrelated values, indicating that the spatial distributions of the $P_0$ and $P_1$ rungs become nearly random. The persistence of the Haldane spin structure in this regime must therefore be encoded in collective correlations among the remaining spin-carrying rungs, rather than being attributable solely to enhanced local hole pairing. This interpretation is supported by the defect-resolved spin correlations shown in Fig.~4(d) of the main text.

\begin{figure}[t]
	\centering
	\includegraphics[width=0.6\linewidth]{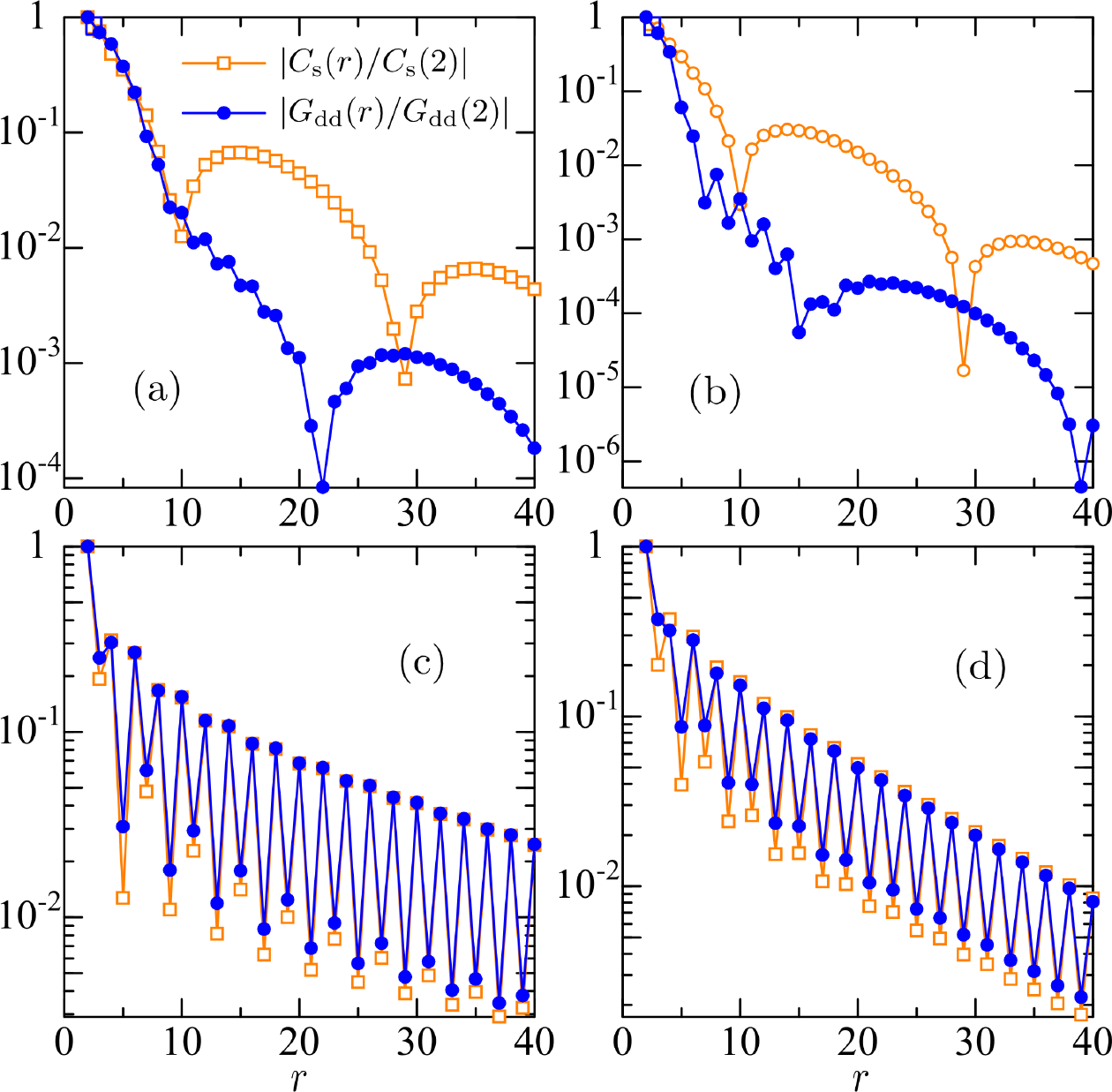}
	\caption{
		Normalized absolute total rung-spin and conditional
		defect-spin correlations at $J=0.4$.
		(a,b) $\delta=0.05$ for $J_{\rm H}=0.5$ and $3.0$,
		respectively.
		(c,d) $\delta=0.5$ for the same respective Hund couplings.
		Orange open symbols show $|C_s(r)/C_s(2)|$,
		and blue filled circles show $|G_{dd}(r)/G_{dd}(2)|$.
		Each correlation function is normalized independently
		at $r=2$. Lines are guides to the eye.
	}
	\label{fig:SM_conditional_defect_corr}
\end{figure}

To examine whether the contrasting defect-spin correlations in
Fig.~4 of the main text originate solely from the probability
of finding singly occupied rungs at both endpoints, we calculate
a conditional defect-spin correlation.
Using the single-occupancy projector
$\hat p_{1,i}=n_{i,1}+n_{i,2}-2n_{i,1}n_{i,2}$
and the corresponding spin contribution
$d_i^z=\hat p_{1,i}\widetilde S_i^z$, we define
\begin{equation}
	G_{dd}(r)
	=
	\frac{
		\overline{\langle d_i^z d_{i+r}^z\rangle}
	}{
		\overline{\langle \hat p_{1,i}\hat p_{1,i+r}\rangle}
	},
	\qquad r\geq1,
	\label{eq:SM_conditional_defect_corr}
\end{equation}
where the overbar denotes an average over rung positions.
For a homogeneous state with vanishing local magnetization,
the connected and unconnected defect-spin correlations coincide,
and Eq.~(\ref{eq:SM_conditional_defect_corr}) becomes
\begin{equation}
	G_{dd}(r)
	=
	\frac{C_{dd}(r)}{P_1^2 g_{P_1}(r)}.
	\label{eq:SM_conditional_defect_uniform}
\end{equation}
This conditioning removes the joint endpoint-occupancy weight;
it does not assume statistical independence of the spin and
charge degrees of freedom.

Figure~\ref{fig:SM_conditional_defect_corr} compares the spatial
profiles after independently normalizing each correlation
function at $r=2$.
At dilute doping, $\delta=0.05$, the normalized $G_{dd}(r)$
differs markedly from the normalized $C_s(r)$ for both Hund
couplings. It is more strongly suppressed at intermediate
distances and does not reproduce the longer-distance modulation
of $C_s(r)$ over the displayed range.
Thus, the difference observed in the main text cannot be
attributed solely to the joint endpoint-occupancy weight.

At $\delta=0.5$, by contrast, the normalized $G_{dd}(r)$ shares
the dominant oscillatory structure and decay envelope of the
normalized $C_s(r)$ for both Hund couplings, although differences
remain near the oscillation minima.
These comparisons concern the normalized spatial profiles,
not equality of the absolute correlation amplitudes.
Together with the finite bulk spin gap, they support the
interpretation that defect spins participate in the correlated
spin sector even when their positional correlations are weak.

\section{Dilute-electron analysis of the ferromagnetic boundaries}
\label{sec:dilute_analysis}

Here we provide the dilute-electron analysis underlying the interpretation of the ferromagnetic region discussed in the main text. We denote the electron density per site by $n=1-\delta$ and consider the limit $n\rightarrow0$. The lower and upper ferromagnetic boundaries at finite density are denoted by $J_{\mathrm H}^{c,-}(n)$ and $J_{\mathrm H}^{c,+}(n)$, respectively, while $J_{\mathrm H}^{c,+}(0)$ denotes the upper boundary in the dilute-electron limit. The weak- and strong-Hund boundaries have different microscopic descriptions, but both reflect the competition among carrier motion, ferromagnetic Hund coupling, and antiferromagnetic leg exchange.

\medskip
\noindent\textit{Weak-Hund boundary: squeezed-chain description.}---
We first consider the weak-Hund side. At $J_{\mathrm H}=0$, the two legs are independent one-dimensional $t$--$J$ chains. In the dilute regime, their charge and spin degrees of freedom can be described, to leading order, using an Ogata--Shiba-type factorized picture~\cite{Ogata1990}, in which the charge sector is represented by spinless fermions, whereas the spins reside on a squeezed chain. Since nearest-neighbor hopping in one dimension does not exchange the order of the particles, the leading kinetic energy is insensitive to their spin configuration.

For spinless fermions of density $n$, the nearest-neighbor density correlator is
\begin{equation}
	C_{nn}(n)
	=
	\left\langle n_i n_{i+1}\right\rangle_{\mathrm{SF}}
	=
	n^2
	-
	\left[
	\frac{\sin(\pi n)}{\pi}
	\right]^2.
	\label{eq:SM_Cnn}
\end{equation}
The leg exchange acts only when two consecutive particles in squeezed space occupy neighboring sites in the original chain. The corresponding conditional contact probability is $C_{nn}(n)/n$, giving the leading squeezed-chain exchange
\begin{equation}
	J_b(n)
	=
	J\frac{C_{nn}(n)}{n}.
	\label{eq:SM_Jb}
\end{equation}

For the exchange convention used in Eq.~(1) of the main text,
\begin{equation}
	J\left(
	\mathbf{S}_i\cdot\mathbf{S}_{i+1}
	-
	\frac14n_i n_{i+1}
	\right),
\end{equation}
the ground-state energy per bond of a fully occupied spin-$1/2$ Heisenberg chain is $-J_b\ln 2$. Since there are $n$ squeezed spins per original site and two legs per rung, the leading antiferromagnetic energy per rung is
\begin{equation}
	e_{\mathrm{AF}}
	\simeq
	-2nJ_b\ln 2
	=
	-2J C_{nn}(n)\ln 2.
	\label{eq:SM_EAF}
\end{equation}

We compare this energy gain with the leading Hund-energy gain of the fully polarized state. At $J_{\mathrm H}=0$, the two legs are independent, so their smooth same-rung occupation probability is, to zeroth order,
\begin{equation}
	\left\langle n_{i,1}n_{i,2}\right\rangle
	\simeq
	n^2.
\end{equation}
The interleg spin correlation vanishes in the nonferromagnetic reference state at this order. For a triplet rung,
\begin{equation}
	\left\langle
	\mathbf{S}_{i,1}\cdot\mathbf{S}_{i,2}
	\right\rangle
	=
	\frac14,
\end{equation}
and the leading Hund energy of the fully polarized state is therefore
\begin{equation}
	e_{\mathrm H}^{\mathrm{FM}}
	\simeq
	-\frac{J_{\mathrm H}}{4}n^2.
	\label{eq:SM_EHundFM}
\end{equation}
The leg-exchange contribution vanishes identically in the fully polarized state because
\begin{equation}
	\mathbf{S}_{i,a}\cdot\mathbf{S}_{i+1,a}
	=
	\frac14n_{i,a}n_{i+1,a}.
\end{equation}
The leading kinetic energies of the competing spin configurations are the same within the squeezed-chain approximation.

Equating the magnitudes of the antiferromagnetic and Hund-energy gains gives
\begin{equation}
	J_{\mathrm H}^{c,-}(n)
	=
	8J\ln 2
	\frac{C_{nn}(n)}{n^2},
\end{equation}
or explicitly,
\begin{equation}
	J_{\mathrm H}^{c,-}(n)
	=
	8J\ln 2
	\left[
	1-
	\left(
	\frac{\sin(\pi n)}{\pi n}
	\right)^2
	\right].
	\label{eq:SM_dilute_lower}
\end{equation}
For $n\ll1$,
\begin{equation}
	\frac{\sin(\pi n)}{\pi n}
	=
	1-\frac{\pi^2n^2}{6}
	+O(n^4),
\end{equation}
which yields
\begin{equation}
	J_{\mathrm H}^{c,-}(n)
	=
	\frac{8\pi^2\ln 2}{3}
	Jn^2
	+
	O(n^4)
	\simeq
	18.24Jn^2.
	\label{eq:SM_dilute_lower_asymp}
\end{equation}
Thus, the weak-Hund ferromagnetic boundary necessarily collapses toward $J_{\mathrm H}=0$ as the electron density vanishes.

Equation~(\ref{eq:SM_dilute_lower}) should be regarded as a leading organizing principle rather than a quantitatively exact transition line. It assumes a factorized charge background and neglects the feedback of finite $J_{\mathrm H}$ on both the charge correlations and the spin state. Nevertheless, it reproduces two robust features of the numerical boundary: its linear dependence on $J$ and its $J_{\mathrm H}^{c,-}(n)\propto n^2$ collapse in the dilute-electron limit.

\medskip
\noindent\textit{Strong-Hund limit: rung-triplet bound pairs.}---
The strong-Hund boundary is controlled by a different low-energy degree of freedom. Consider first two electrons, one on each leg. In the triplet channel, the Hund interaction produces an attractive contact potential
\begin{equation}
	V_T
	=
	-\frac{J_{\mathrm H}}{4}
\end{equation}
when the two electrons occupy the same rung. At total momentum $K=0$, the relative-coordinate problem is a one-dimensional tight-binding problem with hopping amplitude $2t$. Writing the relative-coordinate wave function as
\begin{equation}
	\psi_r
	\propto
	e^{-\kappa|r|},
\end{equation}
the energy away from the contact point is
\begin{equation}
	E
	=
	-4t\cosh\kappa.
\end{equation}
The contact condition gives
\begin{equation}
	\frac{J_{\mathrm H}}{4}
	=
	4t\sinh\kappa,
\end{equation}
and hence
\begin{equation}
	E_{\mathrm{pair}}
	=
	-
	\sqrt{
		\left(
		\frac{J_{\mathrm H}}{4}
		\right)^2
		+
		16t^2
	}.
	\label{eq:SM_Epair}
\end{equation}
Since $E_{\mathrm{pair}}<-4t$ for any $J_{\mathrm H}>0$, an isolated two-electron rung-triplet bound state exists for arbitrarily weak attractive Hund coupling. The existence of this two-particle bound state alone, however, does not imply a fully polarized many-electron ground state.

At sufficiently strong $J_{\mathrm H}$, both the fully polarized and non-fully-polarized states can exploit the same local triplet binding. The leading Hund binding energy therefore ceases to distinguish the competing total-spin sectors. Their energy difference is instead controlled by the motion and mutual spin alignment of the composite triplet pairs.

The motion of a tightly bound triplet pair arises in second order in $t$. Starting from a pair on rung $i$, one electron first hops to rung $i+1$, producing an intermediate configuration in which the triplet pair is broken. At strong coupling, the corresponding energy cost is approximately $J_{\mathrm H}/4$. There are two equivalent hopping sequences, depending on which leg moves first, giving the leading pair-hopping amplitude
\begin{equation}
	t_T
	\simeq
	\frac{8t^2}{J_{\mathrm H}}.
	\label{eq:SM_tT}
\end{equation}
The triplet-pair mobility therefore decreases as $1/J_{\mathrm H}$.

Within the rung-triplet manifold, the individual spin operators satisfy
\begin{equation}
	\mathbf{S}_{i,a}
	\longrightarrow
	\frac12\mathbf{T}_i,
\end{equation}
where $\mathbf{T}_i$ is the spin-$1$ operator of an occupied triplet rung. For two neighboring occupied triplet rungs, the full leg-exchange term, including its density contribution, projects as
\begin{equation}
	J\sum_{a=1}^{2}
	\left(
	\mathbf{S}_{i,a}\cdot\mathbf{S}_{i+1,a}
	-
	\frac14n_{i,a}n_{i+1,a}
	\right)
	\longrightarrow
	\frac{J}{2}
	\left(
	\mathbf{T}_i\cdot\mathbf{T}_{i+1}
	-
	1
	\right).
	\label{eq:SM_triplet_projection}
\end{equation}
The spin-dependent antiferromagnetic interaction between neighboring triplet rungs therefore has the effective coupling
\begin{equation}
	J_{\mathrm{eff}}
	=
	\frac{J}{2}.
	\label{eq:SM_Jeff}
\end{equation}
To leading order, the strong-Hund problem is consequently described by mobile spin-$1$ triplet pairs with kinetic scale $t_T\sim t^2/J_{\mathrm H}$ competing with the antiferromagnetic scale $J_{\mathrm{eff}}=J/2$.

This strong-coupling description is intended for the finite-$J_{\mathrm H}$ regime relevant to the observed upper ferromagnetic boundary. In the strict $J_{\mathrm H}\rightarrow\infty$ limit, $t_T\rightarrow0$, and the charge motion eventually becomes slower than the spin scale. Additional clustering or phase-separation tendencies may then become important. The strict infinite-Hund limit should therefore not be identified with an exact second spin--charge-factorized limit.

\medskip
\noindent\textit{Four-electron problem and the upper ferromagnetic boundary.}---
The minimal system capable of probing the relative spin alignment of two rung-triplet pairs contains four electrons. We therefore study the four-electron problem using finite DMRG with open boundary conditions.

\begin{figure}[t]
	\centering
	\includegraphics[width=0.8\textwidth]{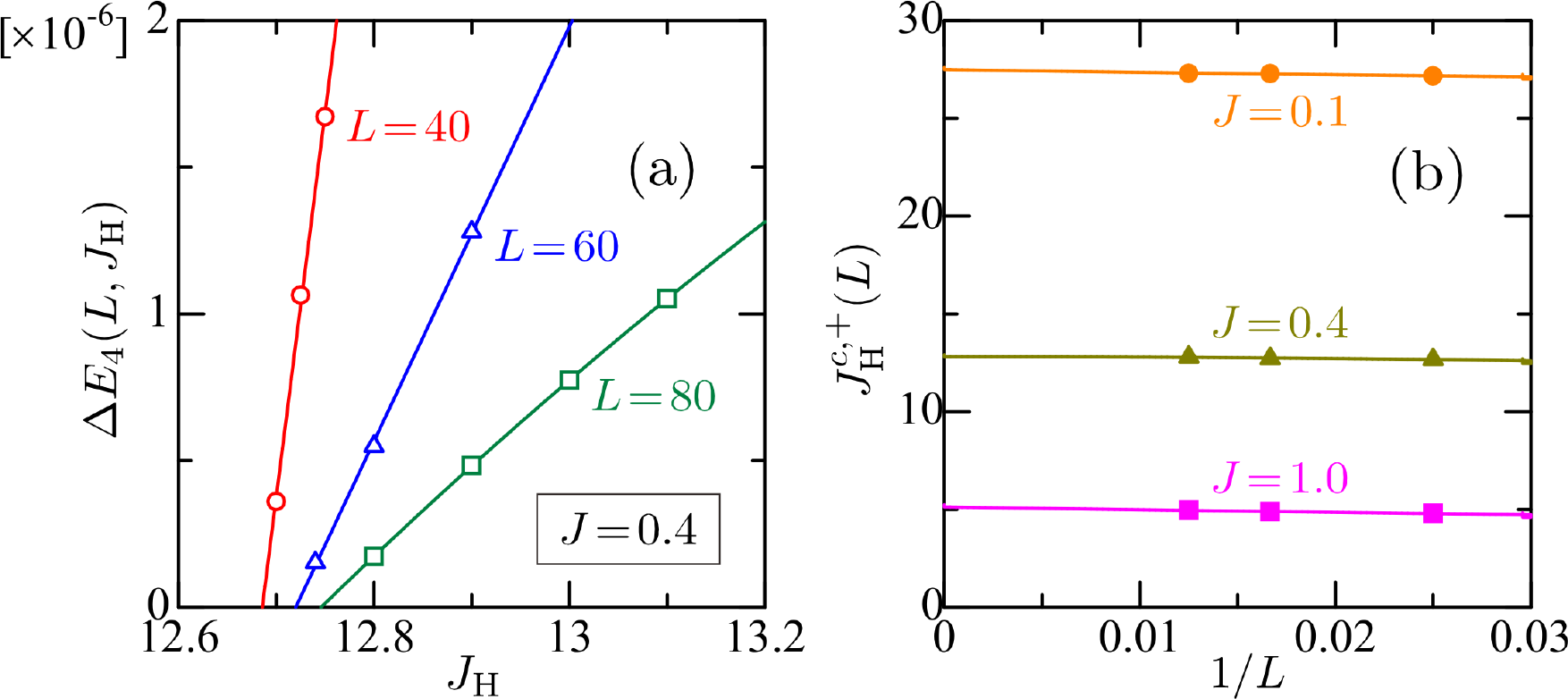}
	\caption{
		Four-electron DMRG estimate of the upper ferromagnetic boundary in the dilute-electron limit.
		(a) Spin-sector energy difference $\Delta E_4(L,J_{\mathrm H})$ near the upper boundary for $J=0.4$ and $L=40$, $60$, and $80$. The solid lines are linear fits to the positive values used to locate the finite-size thresholds $J_{\mathrm H}^{c,+}(L)$.
		(b) Empirical linear extrapolations of $J_{\mathrm H}^{c,+}(L)$ in $1/L$ for $J=0.1$, $0.4$, and $1.0$. The extrapolated values $J_{\mathrm H}^{c,+}(0)/t$ are approximately $27.45$, $12.80$, and $5.14$, respectively. Open boundary conditions are used, and $t=1$.
	}
	\label{fig:four_electron_extrapolation}
\end{figure}

For $N=4$ electrons on a ladder of length $L$, we define
\begin{equation}
	\Delta E_4(L,J_{\mathrm H})
	=
	E_0(L,N=4,S_{\mathrm{tot}}^z=2)
	-
	E_0(L,N=4,S_{\mathrm{tot}}^z=0).
	\label{eq:four_electron_energy}
\end{equation}
In a fully polarized ground state with $S_{\mathrm{tot}}=2$, spin-rotation symmetry requires the $S_{\mathrm{tot}}^z=0$ and $S_{\mathrm{tot}}^z=2$ components of the same multiplet to be degenerate. Consequently, $\Delta E_4=0$. On the strong-Hund side of the ferromagnetic region, a non-fully-polarized state becomes the ground state in the $S_{\mathrm{tot}}^z=0$ sector, and $\Delta E_4$ becomes positive.

For each system size, the finite-size upper critical coupling $J_{\mathrm H}^{c,+}(L)$ is determined by linearly extrapolating the positive values of $\Delta E_4(L,J_{\mathrm H})$ immediately above the boundary to $\Delta E_4=0$. Figure~\ref{fig:four_electron_extrapolation}(a) illustrates this procedure for $J=0.4$ and $L=40$, $60$, and $80$. The DMRG energies were converged until the resulting $\Delta E_4$ was stable on the scale displayed in the figure. This criterion identifies the onset of a finite spin-sector splitting and does not rely on a crossing between two independently tracked energy branches.

At fixed electron number $N=4$, the electron density vanishes as $L\rightarrow\infty$. We therefore use the empirical finite-size form
\begin{equation}
	J_{\mathrm H}^{c,+}(L)
	=
	J_{\mathrm H}^{c,+}(0)
	+
	\frac{a}{L}
	\label{eq:SM_four_electron_scaling}
\end{equation}
to estimate the upper ferromagnetic boundary in the dilute-electron limit,
\begin{equation}
	J_{\mathrm H}^{c,+}(0)
	=
	\lim_{L\rightarrow\infty}
	J_{\mathrm H}^{c,+}(L).
	\label{eq:dilute_upper_boundary}
\end{equation}
As shown in Fig.~\ref{fig:four_electron_extrapolation}(b), the finite-size dependence is weak over the sizes considered. The extrapolated values are
\begin{equation}
	\begin{gathered}
		J_{\mathrm H}^{c,+}(0)/t\simeq27.45
		\quad (J/t=0.1),\\
		J_{\mathrm H}^{c,+}(0)/t\simeq12.80
		\quad (J/t=0.4),\\
		J_{\mathrm H}^{c,+}(0)/t\simeq5.14
		\quad (J/t=1.0).
	\end{gathered}
	\label{eq:upper_critical_couplings}
\end{equation}

These values explain the markedly different appearance of the finite-density phase diagrams in Fig.~3 of the main text. For $J/t=0.1$, the dilute-limit upper boundary lies at $J_{\mathrm H}^{c,+}(0)/t\simeq27.45$, far beyond the range displayed in Fig.~3(b). For $J/t=0.4$, it is reduced to $J_{\mathrm H}^{c,+}(0)/t\simeq12.80$, but remains outside the range displayed in Fig.~3(a). For $J/t=1.0$, however, $J_{\mathrm H}^{c,+}(0)/t\simeq5.14$ falls within the displayed range, consistent with the closing of the ferromagnetic region into the high-doping pocket shown in Fig.~3(c).

Interestingly, the upper-boundary estimates for $J/t=0.4$ and $1.0$ give nearly identical dimensionless products,
\begin{equation}
	\frac{
		JJ_{\mathrm H}^{c,+}(0)
	}{
		t^2
	}
	\simeq
	5.12
	\quad
	(J/t=0.4),
	\qquad
	5.14
	\quad
	(J/t=1.0).
	\label{eq:upper_boundary_products}
\end{equation}
Over this range of $J$, the upper boundary can therefore be summarized empirically as
\begin{equation}
	J_{\mathrm H}^{c,+}(0)
	\simeq
	5.13\frac{t^2}{J}.
	\label{eq:upper_boundary_estimate}
\end{equation}
This relation is consistent with the competition between the decreasing triplet-pair kinetic scale $t_T\sim t^2/J_{\mathrm H}$ and the projected antiferromagnetic scale $J_{\mathrm{eff}}=J/2$. It is, however, only an empirical parametrization over the range $0.4\lesssim J/t\lesssim1$ and should not be extrapolated quantitatively to substantially smaller $J$. In particular, Eq.~(\ref{eq:upper_boundary_estimate}) does not reproduce the result $J_{\mathrm H}^{c,+}(0)/t\simeq27.45$ at $J/t=0.1$.

Using Eqs.~(\ref{eq:SM_tT}) and (\ref{eq:SM_Jeff}), the ratio of the two effective scales at the upper boundary is
\begin{equation}
	\frac{t_T}{J_{\mathrm{eff}}}
	\simeq
	\frac{
		16t^2
	}{
		JJ_{\mathrm H}^{c,+}(0)
	}
	\simeq
	3.12.
	\label{eq:pair_scale_ratio}
\end{equation}
This value applies to both $J/t=0.4$ and $1.0$. Thus, the triplet pairs remain mobile at the observed upper boundary, despite the eventual suppression of their mobility as $J_{\mathrm H}\rightarrow\infty$.

\medskip
\noindent\textit{Kinetic origin of the reentrant ferromagnetic region.}---
The two dilute limits provide a unified interpretation of the reentrant phase structure.

On the weak-Hund side, the one-dimensional charge motion approaches the spin--charge-factorized regime. Because the leading hopping energy is insensitive to the spin configuration, it does not independently stabilize ferromagnetism, whereas the squeezed antiferromagnetic exchange provides the leading spin-dependent energy gain. Its competition with the Hund energy gives $J_{\mathrm H}^{c,-}(n)\propto Jn^2$ and restores the nonferromagnetic phase on the weak-Hund side. Where the numerical spin-gap, string-order, and entanglement-spectrum diagnostics remain finite, this nonferromagnetic phase is identified as the doped Haldane regime. The strictly decoupled point $J_{\mathrm H}=0$, however, should not itself be identified with a Haldane phase.

On the strong-Hund side, electrons form local rung-triplet pairs. The dominant triplet binding is shared by the competing total-spin sectors and therefore no longer provides a selective ferromagnetic energy gain. At the same time, the pair kinetic scale decreases as $t_T\sim t^2/J_{\mathrm H}$, whereas the projected antiferromagnetic scale remains of order $J/2$. Antiferromagnetic correlations consequently recover beyond the upper threshold $J_{\mathrm H}^{c,+}(0)$ inferred from the onset of a positive four-electron spin-sector splitting.

Ferromagnetism is therefore favored only in the intermediate regime in which Hund coupling and carrier motion cooperate most efficiently. This accounts for the reentrant FM pocket within the doped Haldane SPT at high doping and explains why the upper boundary moves rapidly toward larger $J_{\mathrm H}$ as the leg exchange $J$ is reduced.

\bibliography{dopedHaldane}